\documentclass[journal,comsoc,tikz,border=10pt]{IEEEtran}

\usepackage[T1]{fontenc}% optional T1 font encoding

\usepackage{balance}
\usepackage{graphicx}

\ifCLASSINFOpdf
\else
\fi
\usepackage{amssymb,amsmath}
\usepackage[cmintegrals]{newtxmath}
\usepackage{array}
\usepackage{makecell}
\usepackage{multirow}
\usepackage{forest}
\usetikzlibrary{shadows}
\tikzset{every shadow/.style={shadow xshift=5pt,shadow yshift=-5pt}}
\usetikzlibrary{arrows.meta}

\usepackage{cite}

\begin{document}
\title{Data Management in Industry 4.0: \\State of the Art and Open Challenges}

\author{Theofanis~P.~Raptis,
        Andrea~Passarella,
        and~Marco Conti% <-this % stops a space
\thanks{T.~P.~Raptis, A.~Passarella, and M.~Conti are with the Institute of Informatics and Telematics, National Research Council, Pisa 56124, Italy. email: \{t.raptis, a.passarella, m.conti\}@iit.cnr.it.}%
%\thanks{M. Shell was with the Department of Electrical and Computer Engineering, Georgia Institute of Technology, Atlanta, GA, 30332 USA e-mail: (see http://www.michaelshell.org/contact.html).}% <-this % stops a space
\thanks{This work has been partially funded by the European Commission through the FoF-RIA Project AUTOWARE: Wireless Autonomous, Reliable and Resilient Production Operation Architecture for Cognitive Manufacturing (No. 723909).
}%
}

\maketitle

\begin{abstract}
Information and communication technologies are permeating all aspects of industrial and manufacturing systems, expediting the generation of large volumes of industrial data. This article surveys the recent literature on data management as it applies to networked industrial environments and identifies several open research challenges for the future. As a first step, we extract important data properties (volume, variety, traffic, criticality) and identify the corresponding data enabling technologies of diverse fundamental industrial use cases, based on practical applications. Secondly, we provide a detailed outline of recent industrial architectural designs with respect to their data management philosophy (data presence, data coordination, data computation) and the extent of their distributiveness. Then, we conduct a holistic survey of the recent literature from which we derive a taxonomy of the latest advances on industrial data enabling technologies and data centric services, spanning all the way from the field level deep in the physical deployments, up to the cloud and applications level. Finally, motivated by the rich conclusions of this critical analysis, we identify interesting open challenges for future research. The concepts presented in this article thematically cover the largest part of the industrial automation pyramid layers. Our approach is multidisciplinary, as the selected publications were drawn from two fields; the communications, networking and computation field as well as the industrial, manufacturing and automation field. The article can help the readers to deeply understand how data management is currently applied in networked industrial environments, and select interesting open research opportunities to pursue. 
\end{abstract}

% \fi
\IEEEpeerreviewmaketitle

\section{Introduction}

\IEEEPARstart{T}{he} manufacturing industry needs to lead innovations to face the global competitive pressures in the advent of intelligent manufacturing across the broad range of manufacturing sectors \cite{doi:10.1080/0951192X.2016.1258120}. The fourth industrial revolution, or \emph{Industry 4.0} (I4.0), which is being realized in the recent and next years, is expected to deeply change the future manufacturing and production processes, and lead to smart factories and networked industrial environments that will benefit from its main design principles: interoperability, virtualization, decentralization, distributed control and communication, real-time capability, service orientation, quick and easy maintenance, low cost, and modularity \cite{8331177}. In modern industrial applications however, traditional centralized point-to-point control and communication cannot be suitable to meet the increasingly challenging new requirements \cite{7349190}. For this reason, most members of the I4.0 community think in terms of decades rather than years as to when the full I4.0 vision will become state-of-the-art \cite{doi:10.1515/auto-2016-0104}. The I4.0 is highly heterogeneous; in fact it is the aggregation point of more than 30 different fields of the technology \cite{CHIARELLO2018244}.

\begin{figure}[t!]
\centering
\includegraphics[width=0.5\columnwidth]{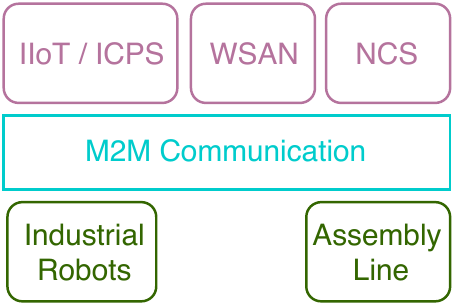}
\caption{Pivotal technological enablers for the I4.0.}
\label{fig::survey40-techs}
\end{figure}

\begin{figure*}[t!]
\centering
\includegraphics[width=\textwidth]{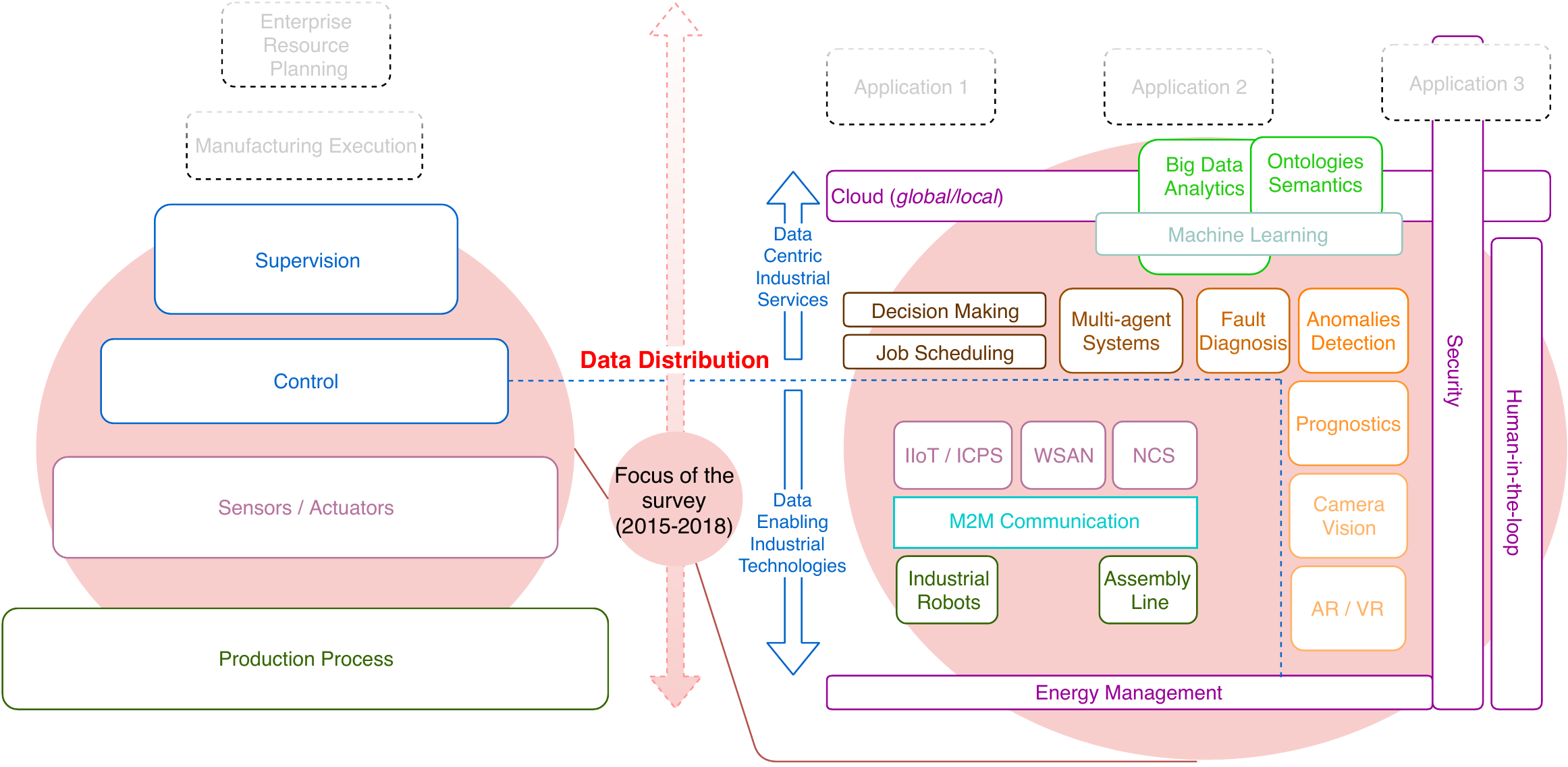}
\caption{Mapping of traditional automation pyramid (left) to I4.0 data enabling technologies and data centric services (right).}
\label{fig::bigpic}
\end{figure*}

In order to address the upcoming challenges of I4.0, several pivotal technological enablers have emerged (Fig.~\ref{fig::survey40-techs}). Novel \emph{assembly lines} used in the production process are expected to boost the reconfiguration of automated manufacturing systems and provide robust operation and short production lifecycles needed by manufacturing firms so as to stay competitive in the marketplace \cite{doi:10.1080/0951192X.2017.1339914}. The \emph{industrial Internet of Things} (IIoT) and the \emph{industrial cyber-physical systems} (ICPS) utilization in industrial settings are expected to revolutionize the way enterprises conduct their business from a holistic viewpoint, i.e., from shop-floor to business interactions, from suppliers to customers, and from design to support across the whole product and service lifecycle \cite{7883993}. The cost decrease coming from \emph{industrial robot} integration in the production process towards mass customization is expected to further improve the robot transparency and promote human-robot collaborations, just as if they were human-human collaborations, since the robot will have ideally the same set of skills and requirements as a human co-worker \cite{8081844}. \emph{Wireless sensor and actuator networks} (WSAN) are able to provide remote monitoring and control of factory plants and machines for the sake of reducing potential equipment failures as well as improving the industrial efficiency and productivity \cite{7892930}. \emph{Networked contol systems} (NCS), which connect cyberspace to physical space enabling the execution of several tasks from long distance, eliminate unnecessary wiring reducing the complexity and the overall cost in designing and implementing industrial solutions \cite{8576512}. The improvements coming from novel customized protocol stacks in machine-to-machine \emph{(M2M) communication}, which achieve multi-gigabyte per-second data rates, submicrosecond latencies, and ultrahigh reliability, are expected to approximate the I4.0 requirements \cite{8048384}.

On top of those technological enablers, groundbreaking services will further boost the I4.0 vision (Fig.~\ref{fig::bigpic}). \emph{Big data analytics}, \emph{machine learning} and \emph{semantic modeling} are expected to make industrial integration easier because the typical data integration involves a lot of data volumes, traffic, mappings and conversions among different data formats \cite{7524742}. \emph{Decision making}, \emph{job scheduling} and \emph{human-in-the-loop} approaches are expected to constitute a kind of hybrid control systems with a dynamic structure and distributed intelligence capable of meeting industrial needs and rapid market changes \cite{TRENTESAUX20161}. \emph{Augmented reality} (AR), \emph{virtual reality} (VR), \emph{camera and vision identification} services are expected to \cite{Berg2017} mimic the human information processing system in order to take advantage of and interpret the ambient industrial environment. \emph{Prognostics} and \emph{prediction} processes, \emph{anomalies detection} and \emph{fault diagnosis} are expected not only to enable the collection of data, but also to support advanced analytics to extract useful insights with high returns on investments in the manufacturing industry \cite{LECHEVALIER201854}. Last but not least, local or global \emph{cloud} integration, smart \emph{energy management} and increased \emph{security} solutions are expected to horizontally fortify a more sustainable production process \cite{7558133}.

\subsection{The crucial role of data}

The natural evolution of those industrial technological enablers and services leads to the generation of huge amounts of \emph{data}; data of many different volumes, traffic and criticality. Data will serve as a fundamental resource to promote I4.0 from machine automation to information automation and then to knowledge automation. In the past several decades, large amounts of data have been generated in the industrial environments, through to the wide use of \emph{networked control systems} (NCS). At the very beginning, those large amounts of data have rarely been used for detailed analyses, which were instead only used for routinely technical checks and process log fulfillments. Later, awareness of the importance in extracting information from data has taken a leading role for the I4.0 \cite{8051033}. This is because there has been an exponential increase in the number of data sources, both archival and in real time. However, data is not equal to value and consequently, to create value with data, one needs data processes which facilitate data reduction to actionable items thus creating value \cite{8334872}.

%In general, manufacturing big data can be divided into three types: device data, product data, and command data.
% ref: A Manufacturing Big Data Solution for Active Preventive Maintenance

\subsection{Contributions of this survey article}

\begin{figure}[t!]
\centering
\includegraphics[width=\columnwidth]{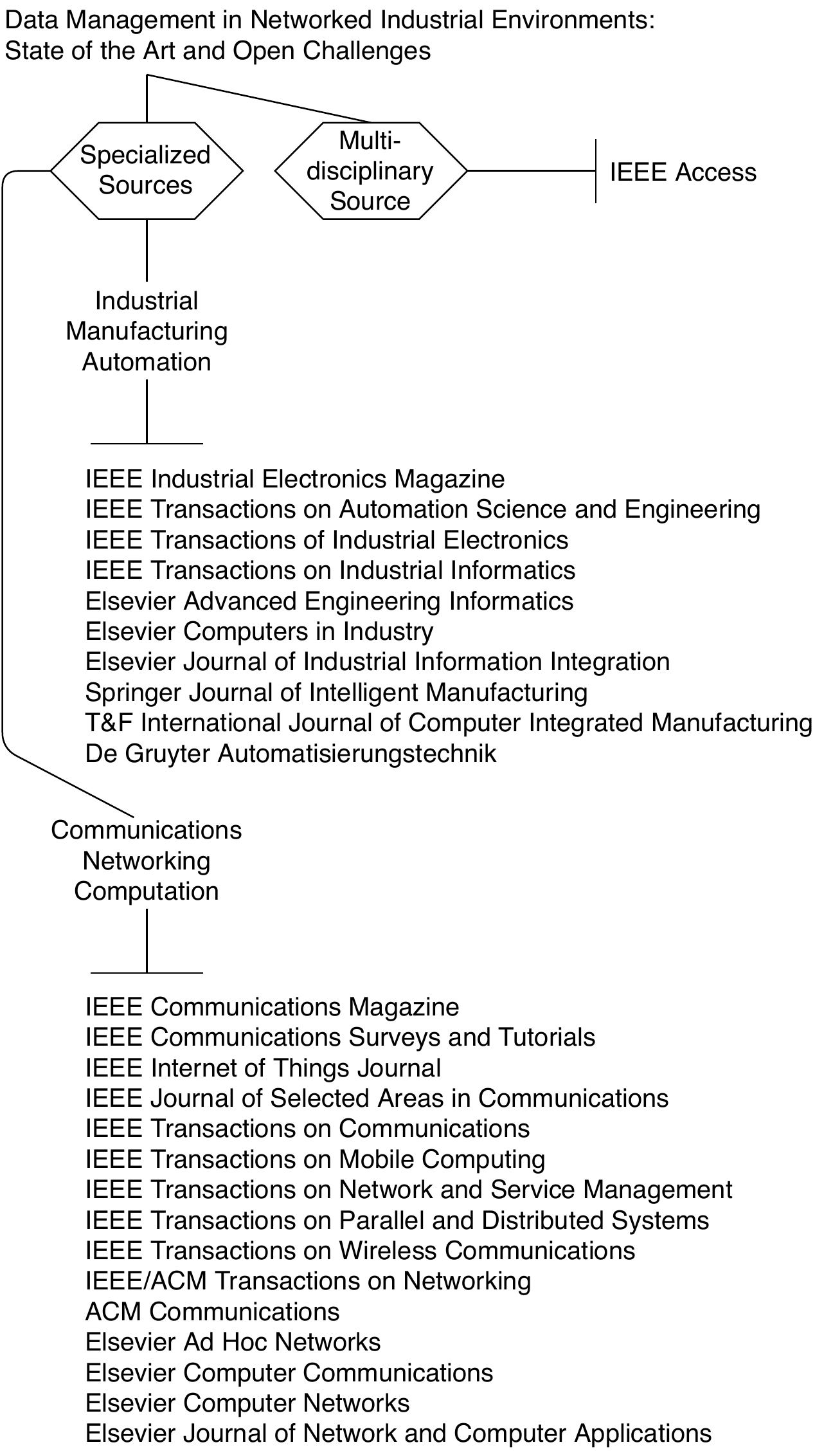}
\caption{Primary sources of information. Focus on two fields: Communications/Networking/Computation and Industrial/Manufacturing/Automation.}
\label{fig::sources}
\end{figure}

This article surveys the literature over the period 2015-2018 on data enabling industrial technologies and data centric industrial services from the point of view of data management as it applies to networked industrial environments and identifies open challenges for the future. A thorough research in two categories of important journals has been conducted, based on two different but complementary groups of scientific fields:

\begin{itemize}
\item Communications, Networking and Computation
\item Industrial, Manufacturing and Automation
\end{itemize}
Fig.~\ref{fig::sources} displays the primary sources of information for this article, identified after an exhaustive literature research. There are some articles coming from some other sources as well, but the list of Fig.~\ref{fig::sources} represents the sources from which the critical mass of the references of this article were drawn. The choice of reported articles is highly selective, due to the fact that in order to be included, an article needs to provide new knowledge on a technological enabler, service, architecture or methodology directly applied on industrial environments. For this reason, a large portion of related literature which investigates similar concepts, but on environments other than industrial, has purposefully been excluded from the current survey.

\begin{figure}[t!]
\centering
\includegraphics[width=\columnwidth]{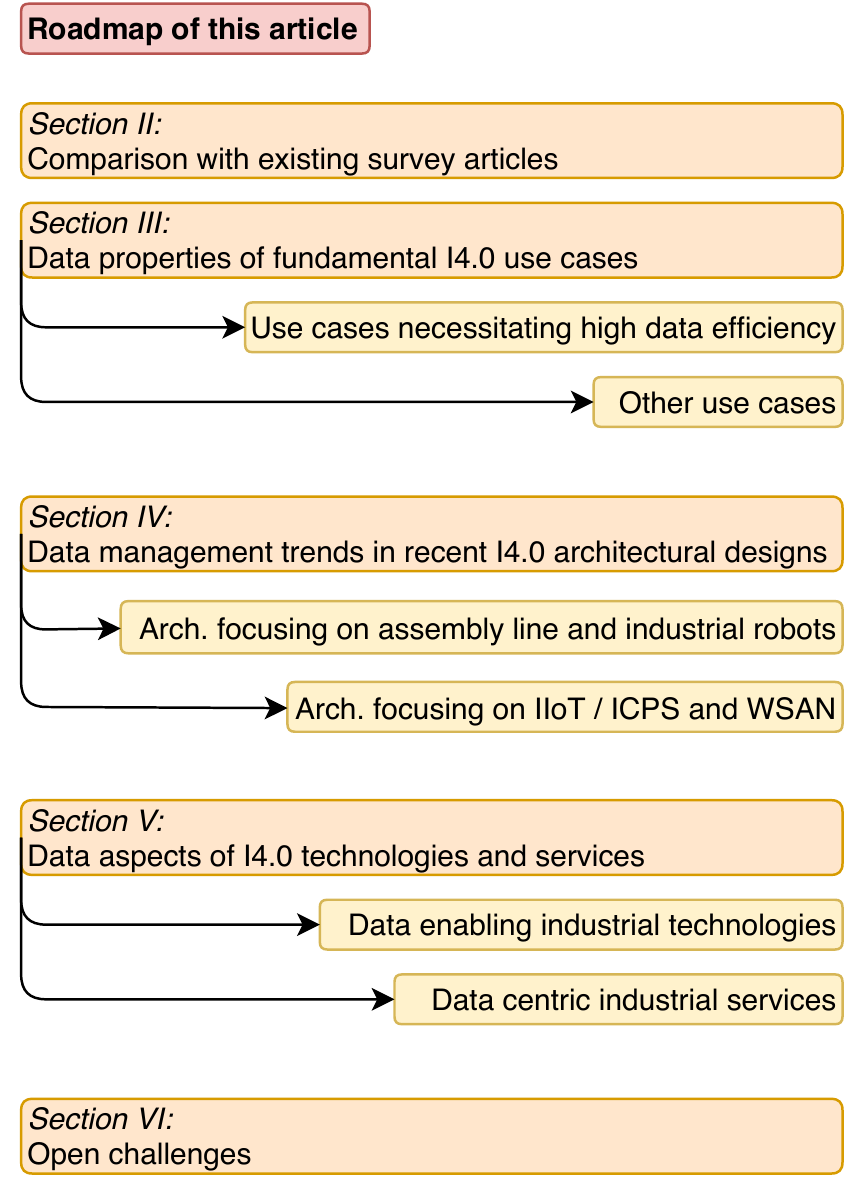}
\caption{Roadmap of this article.}
\label{fig::roadmap}
\end{figure}

Although there are existing surveys which cover some data-centric aspects of  industrial processes, like industrial data management, data-driven manufacturing and cloud manufacturing, to the best of our knowledge, there is no existing survey that covers horizontally, in a holistic way, diverse aspects of data management in heterogenous networked environments of industrial deployments. Consequently, this is the first comprehensive survey which discusses data management in networked industrial environments in a broad view, exposing different use cases, technologies and services that can facilitate the management of distributed data. A comparison to other published surveys is provided in section \ref{sec::sota}. The major contributions of this article are the following. 
\begin{enumerate}
\item An extraction of data properties (volume, variety, traffic, criticality) and an identification of the corresponding data enabling technologies in different I4.0 fundamental use cases, based on practical applications, is provided (section \ref{sec::usecases}). 
\item A detailed outline of recent I4.0 architectural designs with respect to their data management philosophy (data presence, data coordination, data computation) and the extent of their distributiveness (section \ref{sec::architectures}).
\item A holistic survey and taxonomy of the latest I4.0 data enabling technologies (section \ref{sec::techs}) and data centric services (section \ref{sec::services}), spanning all the way from the field level deep in the physical deployments up to the cloud level. This outline is based on an exhaustive research of recent publications and covers the largest part of the I4.0 automation pyramid (Fig.~\ref{fig::bigpic}). 
%\item An in-depth survey of all state-of-the-art of the period 2015-2018 on the following networked I4.0 aspects:
%\begin{itemize}
%\item data enabling technologies (section \ref{sec::techs}).
%\item data centric services (section \ref{sec::services}).
%\end{itemize}
\item A discussion on future interesting open research challenges regarding data management in networked industrial environments (section \ref{sec::open}). 
\end{enumerate}
To the best of our knowledge, such practical survey for data properties, management, technologies and services, for industrial networked environments, coming from recent research contributions does not exist in previous works. The roadmap of this article is displayed in Fig.~\ref{fig::roadmap}.

\section{Comparison with Existing Related Survey Articles} \label{sec::sota}

\begin{table*}[t!]
\begin{center}
\caption{Comparison with existing survey articles on networked industrial environments (2015-2018).}
\label{tab::comparison}
\begin{tabular}{ c | c | c | c || c | c}
\textbf{Articles} & \textbf{Focus area} & \textbf{ \makecell[c]{Focus \\Technologies}} & \textbf{ \makecell[c]{Focus \\Services}} & \textbf{\emph{ \makecell[c]{Data-centric \\aspects}}} & \textbf{\emph{ \makecell[c]{Comments}}} \\\hline\hline
\makecell[c]{Current \\article} && \makecell[c]{IIoT/ICPS, WSAN, \\ Assembly Line,\\Industrial Robots,\\M2M Communication} & \makecell[c]{Machine Learning, \\Multi-agent Systems,\\Big Data Analytics,\\Prognostics, Security\\Human-In-The-Loop\\Energy Management,\\Job Scheduling, \\Decision Making\\Fault Diagnosis,\\Anomalies Detection,\\Ontologies/Semantics\\Camera/Vision/AR/VR} & $\checkmark$ & - \\\hline\hline
\cite{7931543, DIEZOLIVAN201992} &  \makecell[c]{Industrial Data \\Management} & \makecell[c]{IIoT, WSAN, \\ Assembly Line} & \makecell[c]{Machine Learning,\\Big Data Analytics, \\Prognostics,\\Human-In-The-Loop} & $\checkmark$ & \makecell[c]{Smaller number of use cases \\and data centric services} \\\hline
 \makecell[c]{\cite{DEKHTIAR2018227, 6748057, 7775006}} &  \makecell[c]{Data driven \\manufacturing} &  \makecell[c]{IIoT, NCS \\Assembly Line} & \makecell[c]{Machine Learning, \\Multi-agent Systems} & $\checkmark$ & \makecell[c]{Very few use cases and services} \\\hline
 \makecell[c]{\cite{doi:10.1080/0951192X.2013.874595, BABICEANU2016128, doi:10.1080/0951192X.2015.1031704} }&  \makecell[c]{Cloud \\manufacturing} & \makecell[c]{IIoT, WSAN, \\ Assembly Line} & \makecell[c]{Big Data Analytics, \\Job Scheduling,\\Decision Making}& $\checkmark$ & Narrow focus on cloud level \\\hline
 \makecell[c]{\cite{7448365, DEGUGLIELMO20161, TRAPPEY2017208, TELESHERMETO201784}} &  \makecell[c]{Industrial wireless \\standards} & \makecell[c]{IIoT, WSAN,\\M2M Communication} & - & - & \makecell[c]{Narrow focus on \\wireless communication}\\\hline
\makecell[c]{\cite{7892976, 8057758, OESTERREICH2016121, 7104117, 7296649, 8558534, QUEIROZ201796}} &  \makecell[c]{IIoT \\technologies} & IIoT, WSAN & \makecell[c]{Energy Management,\\Security\\Machine Learning} & - & \makecell[c]{Narrow focus on IIoT \\and WSAN technologies} \\\hline
\cite{7299315, OYEWOBI2017140} &  \makecell[c]{Industrial \\cognitive radio}  & \makecell[c]{WSAN,\\M2M Communication} & \makecell[c]{Energy Management,\\Security}  & - & Highly specialized topic \\\hline
 \makecell[c]{\cite{Behnamian2016, DALLASEGA2018205, 7339673}} &  \makecell[c]{Scheduling,\\synchronization} & NCS, Assembly Line & \makecell[c]{Job Scheduling, \\Decision Making} & - & \makecell[c]{Narrow focus on \\scheduling services} \\\hline
 \makecell[c]{\cite{SONG201711, HUANG20151, LYU20171}}&  \makecell[c]{Product \\systems} & Assembly Line & Decision Making & - & \makecell[c]{Narrow focus on \\high-level applications}  \\\hline
\end{tabular}
\end{center}
\end{table*}

The purpose of this article is to provide a holistic overview on data management as it applies to networked industrial environments. Although both data management and industrial networks are quite vibrant research fields, they are rarely mentioned together in a holistic manner. To the best of our knowledge, this is the first time that the topics of data management on the industrial networking realm are systematically extracted, dissected, categorized and put together in a survey article, hence bridging the gap between these two seemingly disconnected yet highly complementary paradigms. There exist, however, several published works that cover in depth multiple niche areas found in our survey. In fact, some of them explore several data centric aspects, but for focused application areas, services and technologies. This section will provide an overview of some of those relevant studies. Table \ref{tab::comparison} displays the comparison with other survey articles focusing on networked industrial environments.

\subsection{Industrial data management} 

The most relevant to this article surveys investigate industrial data management. In \cite{7931543}, the authors present a survey on the IIoT aspects of large-scale petrochemical plants as well as recent activities in communication standards for the IoT in industries, with a slight flavor of data management. The article addresses the key enabling middleware approaches, e.g., and highlights the research issues of data management in the IoT for large-scale petrochemical plants. As such, it is entirely focused on this specific use case. In \cite{DIEZOLIVAN201992}, the authors provide a survey of the recent developments in data fusion and machine learning for industrial prognosis. To this end, a principled categorization of feature extraction techniques and machine learning methods is provided. This analysis is highly focused on the data centric services of machine learning, data fusion and prognostics. Different from those works, we investigate data management aspects in a much wider spectrum of use cases and data centric services.

\subsection{Data-driven manufacturing}

Another group of relevant articles is the surveys investigating data-driven manufacturing. In \cite{DEKHTIAR2018227}, the authors focus on highlighting the major specificities of data engineering and the data-processing difficulties which are inherent to data coming from the manufacturing industry. They specifically emphasize on the data centric services of machine learning and deep learning and consequently the survey is highly focused both in terms of use case and in terms of services. In \cite{6748057}, the authors aim to provide an overview of data-based techniques with recent developments focused on the industrial closed-loop applications like process monitoring and control. Another overview on the model-based control and data-driven control methods is presented in \cite{7775006}. Those two articles focus entirely on control related issues.

\subsection{Cloud manufacturing} 

In \cite{doi:10.1080/0951192X.2013.874595} and \cite{doi:10.1080/0951192X.2015.1031704}, the authors survey the state of the art in the area of cloud manufacturing, identify recent concepts, implementations and technologies, and discuss potential research trends and opportunities. In \cite{BABICEANU2016128}, the authors provide a review of the more specific field of virtualization and cloud-based services for manufacturing systems and of the use of big data analytics for planning and control of manufacturing operations. Although those surveys incorporate some data related concepts, they focus their investigation on the cloud layer of networked manufacturing environments and explore a specific subset of related technologies and services.

\subsection{Industrial wireless standards}

As wireless technologies penetrate more and more the manufacturing landscape, industrial wireless standards are emerging. \cite{7448365} discusses key aspects of the four most popular industrial wireless sensor network standards: ZigBee, WirelessHART, ISA100.11a, and WIA-PA. The detailed design and protocol architectures are comparatively examined. \cite{DEGUGLIELMO20161} provides a clear and structured overview of all the new 802.15.4e mechanisms and describes the details of the main 802.15.4e MAC behavior modes, namely Time Slotted Channel Hopping (TSCH), Deterministic and Synchronous Multi-channel Extension (DSME), and Low Latency Deterministic Network (LLDN). \cite{TRAPPEY2017208} depicts a systematic approach to review IIoT technology standards and patents. The literature of emerging IIoT standards from the International Organization for Standardization (ISO), the International Electrotechnical Commission (IEC) and the Guobiao standards (GB), and global patents issued in US, Europe, China and World Intellectual Property Organization (WIPO) are systematically presented in this study. \cite{TELESHERMETO201784} reviews the scheduling mechanisms for 802.15.4-TSCH and slow channel hopping MAC in low power industrial wireless networks. It categorizes the numerous existing solutions according to their objectives (e.g. high-reliability, mobility support) and approaches and identifies some open challenges, expected to attract much attention over the next few years. All those studies provide an interesting glimpse into the standardization domain for industrial networked environments, but, naturally, their focus is highly specific and is very different from the holistic approach focusing on data management which is presented in our survey.

\subsection{IIoT technologies} 

Due to the fact that IIoT is a core technological enabler for the realization of I4.0, there is a significant number of surveys that report on various IIoT aspects. \cite{7892976} provides an overview of the Industrial Internet with the emphasis on the architecture, enabling technologies, applications, and existing challenges. More specifically, it investigates the enabling technologies of each layer that cover from industrial networking, industrial intelligent sensing, cloud computing, big data, smart control, and security management. Moreover, it discusses the application domains that are gradually transformed by the Industrial Internet technologies, including energy, health care, manufacturing, public section, and transportation. A detailed discussion on design objectives, challenges, and solutions, for WSANs, are presented in \cite{8057758}. A careful evaluation of industrial systems, deadlines, and possible hazards in industrial atmosphere are discussed. The primary objective of \cite{OESTERREICH2016121} is to explore the state of the art as well as the state of practice of  I4.0 relating technologies in the construction industry by pointing out the political, economic, social, technological, environmental and legal implications of its adoption. The recent advancements in FPGA technology, emphasizing the novel features that may significantly contribute to the development of more efficient digital systems for industrial applications are presented in \cite{7104117}.Various proposed controllers for high-mix semiconductor manufacturing processes are surveyed in \cite{7296649} from an application and theoretical point of view. Remaining challenges and directions for future work are also summarized with the intent of drawing attention to these problems in the systems and process control communities. In \cite{8558534}, a comprehensive survey of IIoT technologies has been presented, including IIoT architectural approaches, applications and characteristics, existing research efforts on control, networking and computing systems in IIoT, as well as challenges and future research needs. Finally, in \cite{QUEIROZ201796}, the authors provide an overview of the standards used to implement industrial WSANs and discuss the characteristics of the wireless channel in industrial environments. Different to the current survey, all those articles have an exclusive focus on a subset of technological enablers, IIoT and WSAN technologies.

\subsection{Industrial cognitive radio} 

This is a specialized group of survey articles, which we report in order to provide a complete list of relevant existing survey articles. The relevance to data management is minimal, but, nevertheless, the core technological enabler is already applied to industrial networked environments. \cite{7299315} summarizes cognitive radio methods relevant to industrial applications, covering cognitive radio architecture, spectrum access and interference management, spectrum sensing, dynamic spectrum access, game theory, and cognitive radio network security. \cite{OYEWOBI2017140} highlights and discusses important QoS requirements of IWSN as well as efforts of existing IWSN standards to address the challenges. It also discusses the potential and how cognitive radio and spectrum handoff can be useful in the attempt to provide real-time reliable and smooth communication for IWSNs.

\subsection{Scheduling and synchronization}

An interesting higher level application for the I4.0 is the scheduling and synchronization of multiple factories. \cite{Behnamian2016} provides a review on the multi-factory machine scheduling. It classifies and reviews the literature according to shop environments, including single machine, parallel machines, flowshop, job shop, and open shop. The concept of proximity is used to analyze synchronization between suppliers and the construction site. \cite{DALLASEGA2018205} presents a framework for explaining I4.0 concepts that increase or reduce proximity. The authors find that Industry 4.0 technologies mainly influence technological, organizational, geographical and cognitive proximity dimensions. \cite{7339673} gives a review on recent advances on the analysis and design of fuzzy-model-based nonlinear NCS with various network-induced limitations such as packet dropouts, time delays, and signal quantization. With these network-induced constraints, the developments on various control and filtering design issues are surveyed in details, and some essential technical difficulties are mentioned. 

\subsection{Product-service systems} 

Product-service systems are business models that provide for cohesive delivery of products and services. Product-service system models are emerging as a means to enable collaborative production and consumption of both products and services, with the aim of pro-environmental outcomes \cite{PISCICELLI201521}. They are thus an important application on the top of the I4.0 automation pyramid. \cite{SONG201711} is dedicated to the systematic status survey on product-service systems requirement management. The results of this work provides references for future research in the area of product-service systems development, with the aim of offering integrated and holistic requirements management for product-service systems. It analyzes the state of the art of requirements management for product-service systems by reviewing extensive literature of requirement identification, analysis, specification, and forecast.  \cite{HUANG20151} reviews multiple defect types of various inspected products which can be referenced for further implementations and improvements. The objective of \cite{LYU20171} is to provide a comprehensive literature review on recent research and development in product modeling from three perspectives: product knowledge in product representation, distributed computing in information technology, and product lifecycle in product development process. Contrary to our survey, this group of articles is distant both from data management and from industrial networking technologies. However, it is worthy having it reported, as it is a nice example of I4.0 post-production applications.

In summary, our survey attempts to give a holistic review of the state-of-the-art regarding data management as it applies to networked industrial environments. The review is centered around a plethora of technologies and services brought forth by the relevant I4.0 use cases and architectural designs, and provides a more recent view of the industrial data management field. Our article is an ambitious effort to capture the interplay between data management and networked industrial environments, instead of delving into one particular data centric service or one data enabling technology exclusively. The motivation behind this survey is to provide researchers coming from both the communications/networking/computation fields and the industrial/manufacturing/automation fields a glance of the intersection between these two domains at a higher level.

\section{Data Properties of Fundamental I4.0 Use Cases} \label{sec::usecases}

In this section, we provide a thorough extraction of data properties in different I4.0 fundamental use cases, based on practical applications reported in recent research contributions. To the best of our knowledge, such practical extraction, coming from real world applications and reports does not exist in previous work for the reported activity period. At the same time, we identify the basic set of technological enablers that are needed for the realization of those important use cases, and we use them as a compass for the follow-up analysis which is presented in section \ref{sec::outline}. The extracted data properties about the use cases are displayed in Table \ref{tab::usecases}. Our interest is to extract three specific data properties, in order to understand the data requirements in recent I4.0 use cases. The four data properties we focus on are the following:
\begin{enumerate}
\item \emph{Data volume}: The size of the data to be circulated in a network environment is of crucial importance to the network design and the technological enablers used in the deployment. In industrial networked environments there can be a diversity of data volumes, depending on the scope of each use case. We label as data of \emph{small} volume the data of lower sizes, such as sensor measurements, of \emph{medium} volume the data of higher sizes, such as images or sound files, and of \emph{big} volume, the data of the highest sizes, such as videos and detailed 3D representations.
\item \emph{Data variety}: The diversity of the data can also be variable, according to the use case. We label as \emph{diverse} the data variety in use cases where different kinds of data are needed and as \emph{uniform} the data variety in use cases where similar kinds of data are needed. The data variety can significantly affect algorithmic decisions and service provisioning when targeting efficient solutions per use case.
\item \emph{Data traffic}: Different data varieties, as well as different data generation velocities and use case requirements can lead to diverse traffic patterns in an industrial networked environment. Although deterministic solutions for traffic regulation have started becoming mature for various types of wired industrial deployments, the wireless part is still facing great challenges and comes hand in hand with strict I4.0 requirements. Communication support for industrial automation is challenging in wireless environments as the lossy nature of radio links and node unreliability greatly affects the performance of real-time data delivery. We label as \emph{intense} the data traffic in a network where large amounts of data have to be generated and delivered in small amounts of time, in many cases without predefined global schedules, typically leading to various networking problems necessitating algorithmic solutions for traffic management. On the other hand, we label as \emph{mild} the data traffic in a network where data can be circulated without serious problematic phenomena.
\item \emph{Data criticality}: Data that are not managed according to the underlying I4.0 requirements may adversely affect the performance of system monitoring, control and safety. For example in chemical plant, the chemical leakage must be informed in predefined times \cite{7552640}. This inherent importance separates the data in two major categories, critical and non-critical data. We label the first category as data of \emph{high} criticality and the second category as data of \emph{low} criticality. 
\end{enumerate}
Based on the extracted data properties, we differentiate the use cases in two categories: on the one hand we have the use cases which necessitate a combination of multiple ``heavy'' accomplishments in terms of data requirements and on the other hand we have the use cases with ``light'' data properties. The most important industrial use cases that we identified in the recent literature are the following.

\subsection{Use cases necessitating high data efficiency}

\begin{table*}[t!]
\begin{center}
\caption{Data properties extracted from recent works on various I4.0 use cases.}
\label{tab::usecases}
\begin{tabular}{ r || c | c || c | c | c | c | c }
 \multicolumn{1}{c}{} & \multicolumn{1}{c}{} & \multicolumn{1}{c}{}  & \multicolumn{4}{c}{\emph{\textbf{Data}}}\\\cline{4-7}
\textbf{Use Case} & \textbf{References} & \textbf{Enabling Technologies} & \emph{\textbf{Volume}} & \emph{\textbf{Variety}} & \emph{\textbf{Traffic}} & \emph{\textbf{Criticality}}  \\\hline\hline
Oil / Gas & \cite{7931543, AALSALEM201887, 7423729, 7329985} & \makecell[c]{IIoT, WSAN, \\M2M Communication} & small & uniform & intense & low / high \\\hline
Automotive & \cite{doi:10.1080/0951192X.2017.1356473, 7402269, doi:10.1515/auto-2016-0069} & \makecell[c]{IIoT / ICPS, Assembly Line, \\NCS, Industrial Robots} & small / big & diverse & mild & low \\\hline
%Aeronautics & \cite{7402269, 7921798} & \makecell[c]{IIoT, Assembly Line, \\ WSAN, Industrial Robots} &&& \\\hline
\makecell[r]{Marine \\Vessels} & \cite{doi:10.1080/0951192X.2017.1407452, 6797871, Krishnamurthy:2005:DDI:1098918.1098926, 8281493} &  \makecell[c]{Assembly Line, \\ NCS, Industrial Robots} & small / big & diverse & mild & low / high \\\hline
\makecell[r]{Asset \\Tracking} & \cite{PEASE201798, SATYAVOLU201622, LYLYYRJANAINEN201682, HANSEN2018145} & IIoT / ICPS & small / big & uniform & mild & low \\\hline
%Gears & \cite{6774885, Deng2018} & \makecell[c]{Assembly Line, \\Industrial Robots} &&& \\\hline
\makecell[r]{Customized \\Assembly} & \cite{8291114, doi:10.1080/0951192X.2014.964323} & \makecell[c]{IIoT, Assembly Line, \\NCS, Industrial Robots} & small / big & diverse & intense & high \\\hline\hline %8289327
%Construction & \cite{Kirkpatrick:2018:CIS:3190347.3178312} & IIoT, WSAN &&& \\\hline
\makecell[r]{Crane \\Scheduling} & \cite{HE201559, HE20152464} & IIoT / ICPS & small & uniform & mild & low \\\hline
\makecell[r]{Refrigerated \\Warehouses} & \cite{7104115} & WSAN & small & uniform & mild & low \\\hline
%\makecell[r]{Livestock \\Identification} & \cite{HANSEN2018145} & IIoT &&& \\\hline
\makecell[r]{Healthcare \\Monitoring} & \cite{7738500} & WSAN & small & uniform & mild & low \\\hline
\makecell[r]{Production \\Control} & \cite{7862844, 7572893, 7342942} & \makecell[c]{IIoT, NCS, \\Assembly Line} & small / big & diverse & mild / intense & low \\\cline{4-7}
\end{tabular}
\end{center}
\end{table*}

\subsubsection{Oil / Gas} 

Large-scale petrochemical plants %(Fig.~\ref{fig::helpe}) 
incorporate dense wireless devices such as RFID tags for machine identification, sensors for large-scale rotational equipment monitoring and fault diagnosis, and employ IIoT technologies for tight and seamless integration between lower layer components, such as sensors and actuators, to the higher level connected with the cloud platforms \cite{7931543}. In order to ensure the safety of production sites in large petrochemical industries  \cite{7423729}, and long interconnected gas networks \cite{7329985} those sensorial artifacts are positioned around gas pipes, targeting 24/7 monitoring. Data generated by the wireless sensors about parameters and abnormal events are processed for decision making thereby improving production, predicting maintenance and failures for the industrial equipment. Data usually come from sensor devices in small volumes, typically including sensor measurements of various types. Although the variety can be limited to the various sensor readings, there can be increased wireless traffic in the network; a result of thousands of sensors operating simultaneously both in real-time and periodically. The use case offers a mix of both critical and non-critical applications. An example of the first is a gas leakage must be informed as soon as possible. An example of the second is the predictive maintenance of a set of gas pipes over an interval of some years.

%\begin{figure}[h!]
%\centering
%\includegraphics[width=\columnwidth]{helpe}
%\caption{Large-scale petrochemical industrial deployment in Aspropyrgos, Greece, comprised of petrochemical materials, metal pipes, fuel stations, oil refining equipment and factory units (\copyright Hellenic Petroleum S.A.)}
%\label{fig::helpe}
%\end{figure}

\subsubsection{Automotive} 

In the last two decades, distributed embedded electronic applications have become the norm in a large part of the automotive assembly industry. %(Fig.~\ref{fig::fiat}) 
Due to critical requirements and the distributed nature of the various ECUs implementing assembly functions, the validation of end-to-end timing constraints on those networked industrial environments has become an important part of the design process of a car \cite{7402269}. In addition to existing stand-alone solutions, cooperating networked information and control systems are increasingly used as tools for the coordination of this challenge for production support \cite{doi:10.1515/auto-2016-0069}. The volume of generated data can vary in the automotive production process, providing also a great range of diversity. For example, there can be small volumes of data (positioning systems with various sensors for determination of the exact position of vehicles, tools, resources and processes), as well as big volumes of data (assembly assistance system through monitors or data glasses which guide the workers during their working process, by exploiting audio-visual data). The majority of the generated data is usually distributed via wired deterministic networks, and for this reason the traffic can be regulated in an offline, centralized manner. For the same reason, the data criticality is not significantly high in this type of use case.

%\begin{figure}[h!]
%\centering
%\includegraphics[width=\columnwidth]{fiat}
%\caption{Mirafiori assembly plant, Turin, Italy, comprised of an assembly line, cyber-physical components and human operators (\copyright Fiat Chrysler Automobiles N.V.)}
%\label{fig::fiat}
%\end{figure}

%\emph{Aeronautics}: 

\subsubsection{Marine Vessels}
 
Today's shipbuilding industry %(Fig.~\ref{fig::elefsis}) 
is characterized by one-off manufacturing and complex construction processes, and as such, it is difficult to estimate a construction process at the planning stage and many diverse problems are involved, such as backorders and over-loaded capacity between consecutive processes \cite{doi:10.1080/0951192X.2017.1407452}. Data processing, can be used for fault detection and diagnosis in such complex industrial processes, starting from the construction stage of a marine vessel and finishing at its running operation \cite{6797871}. Sensing technology is a cornerstone for many industrial applications, including preventative equipment maintenance, both inside fabrication plants and onboard the marine vessels \cite{Krishnamurthy:2005:DDI:1098918.1098926}. Recent shipbuilding industry advancements introduce production management methodologies and a pre-verification in virtual environments. Related tools facilitate the traffic and criticality constraints on the production phase and lower their intensity \cite{8281493}. Similar to the automotive industry, the volume of generated data can vary in the marine vessel production process, providing also a great range of diversity.

%\begin{figure}[h!]
%\centering
%\includegraphics[width=\columnwidth]{elefsis}
%\caption{Elefsis Shipyards, Elefsina, Greece comprised of cranes and dry docks (\copyright Kathimerini Publishing S.A.)}
%\label{fig::elefsis}
%\end{figure}

\subsubsection{Asset Tracking}

Mass production in manufacturing puts greater emphasis on real-time asset location monitoring %(Fig.~\ref{fig::asset}), 
which renders the sensor data to be of paramount importance. When location information can be associated with monitored contextual information, e.g. machine power usage and vibration, it can be used to provide smart monitoring information, such as which components have been machined by a worn or damaged tool \cite{PEASE201798}. RFID is the most commonly utilized product tracking and automation technology, especially useful in the supply chain industry \cite{SATYAVOLU201622}, as well as in more specialized industries of asset tracking like identification of individual farm animals \cite{HANSEN2018145}. The generated data can be diverse over all asset tracking applications, but usually only one tracking method is used for each individual application, leading to a uniform data variety. The volume of the data also varies per application, coming from some simple RFID readings in product tracking to images or videos in farm identification. The data criticality is low, as the related data processing and calculations are conducted a posteriori.

%\begin{figure}[h!]
%\centering
%\includegraphics[width=\columnwidth]{asset}
%\caption{RFID asset tracking and management in the warehouse, with identification, assignment, location and movement records (\copyright RFID4U - eSmart Source, Inc.)}
%\label{fig::asset}
%\end{figure}

\subsubsection{Customized Assembly}

Serial assembly lines are mainly used for large scale production since they can provide short cycle times and high production rates with high efficiency in terms of cost, time and quality. In pursuit of flexibility, different paradigms have been investigated in terms of automation level and production system organization \cite{doi:10.1080/0951192X.2014.964323}, like customized assembly lines. %(Fig.~\ref{fig::sfkl}). 
IIoT integrates the key technologies of industrial communication, computing, and control so as to provide a new way for a wide range of assembly resources to optimize management and dynamic scheduling \cite{8291114}. With the technological enablers on flexible assembly lines ranging from IIoT and ICPS to robotic bimanipulators, NCS and moving robots, it is natural that there is a great diversity of data resources to be analyzed. The volumes of data significantly differ from application to application. For example, in the case of mobile robotic assembly, large volumes of motion data are usually exchanged between the different controllers for further data fusion, while in the case of custom part identification, smaller identification data are needed. This use case family is usually characterized by a high criticality factor, due to the fact that the assembly process has to be quick and accurate, affecting accordingly the related data processes.

%\begin{figure}[h!]
%\centering
%\includegraphics[width=\columnwidth]{sfkl}
%\caption{Customized assembly lines at SmartFactory$^\text{KL}$, Germany (\copyright SmartFactory$^\text{KL}$)}
%\label{fig::sfkl}
%\end{figure}

\subsection{Other usecases}

%\emph{Gears} The design and manufacturing of gears is still a hot topic of research that is vital for application in helicopter transmissions, motorcycle gears, reducers, and in other branches of industry. Meanwhile, the growing demands of application in industry and market for high-quality spiral bevel gears make these improvements urgent. Consequently, state-of-the-art machinery and techniques in manufacturing spiral bevel gears are being explored \cite{Deng2018}.

%\emph{Construction}

\subsubsection{Crane Scheduling}

Container terminals %(Fig.~\ref{fig::crane}) 
have to improve their service efficiency to seek the optimal trade-off between energy-saving and service efficiency improvement. Since the energy consumption and service efficiency of container terminals are mainly contributed by the handling cranes, the scheduling of the handling cranes is critical \cite{HE201559}. Moreover, with the increase of sizes of container vessels, container terminals are encountering another challenge, i.e., the rapid handling of containers for mega-vessels. Thus, container terminals must shorten the vessel turnaround time, which is an influential factor of their service level \cite{HE20152464}. Due to the fact that the necessary computations are conducted in an offline manner, usually via optimization modules, the data properties of this use case are simple. An input module, which is the basis for generating crane schedules and evaluating the schedules, consists of two data parts: static data and dynamic data. The static data part include all parameters such as the handling volume of each container, the time window on each container and the handling
efficiency of each crane. The other parameters are used for evaluation, such as the cost of unit energy consumption. The dynamic data include all decision variables, which are generated by the optimization module.

%\begin{figure}[h!]
%\centering
%\includegraphics[width=\columnwidth]{crane}
%\caption{Crane scheduling at the Port of Genoa, Italy, including container terminals and yard cranes (\copyright Western Ligurian Sea Port Authority)}
%\label{fig::crane}
%\end{figure}

\subsubsection{Refridgerated Warehouses}

Changing the cold storage temperature set points of the refrigerated warehouses will cause the reduction of product quality and further increase economic costs to the industrial consumers. Reduction of the electricity price on the grid, the total costs of maintenance, and the total energy consumption comparing has recently been a target objective of operations research \cite{7104115}. This use case is characterized by small volumes of sensor data (mainly temperature), periodically sent to a central control station for long term planning.

\subsubsection{Healthcare monitoring}

Industrial manufacturing has recently started embedding new functions in the form of safety monitoring or smart factories. Another recent trend of interest is the combination of heterogeneous services from different fields for providing automated healthcare services in industrial environments \cite{7738500}. As with typical monitoring use cases, the data come in small volumes, from a range of different but limited sensors targeting long term or real-time healthcare optimization.

\subsubsection{Production Control}

Controlling the various stages and processes during the production process has attracted a widespread research interest in various areas, ranging from the shop floor with vibration control \cite{7862844}, PLC design control \cite{7572893} up to the application layer with economic optimizations \cite{7342942}. Depending on the layer of the industrial integration we are considering, data volumes can be small or large, and the related traffic in the networked environment low or high. 

\section{Data Management Trends in Recent I4.0 Architectural Designs} \label{sec::architectures}

In this section we attempt to place recent architectural innovations in the broader context of networked industrial environments by surveying the fundamentals of both recently proposed I4.0 enabling architectures and by extracting the data management philosophy of these architectural alternatives. The section's primary emphasis concerns data related concepts, rather than specific architectural constructs. A number of research teams have proposed the development of relevant architectures which incorporate either directly or indirectly some kind of data management interfaces and control mechanisms across one or more architectural layers. For the reported period, 2015-2018, the most important I4.0 enabling architectural designs have been presented in \cite{8241718, BONEV201558, 7883994, doi:10.1080/0951192X.2016.1185155, SADOK201558, 7785890, doi:10.1080/0951192X.2015.1066861, 7498096, doi:10.1080/0951192X.2014.1003411, 7466847, electronics7120400, XUE201628, doi:10.1080/0951192X.2017.1392615, 8295235, 7750612, CAMPANELLI201547, 7885069, 7358111, 7152942, doi:10.1080/0951192X.2015.1130242, doi:10.1515/auto-2015-0017, DELARAM201896, doi:10.1080/0951192X.2015.1125766, doi:10.1080/0951192X.2015.1032355, 8333734}.

\begin{table*}[h!]
\begin{center}
\caption{Data management trends in recent I4.0 architectural designs.}
\label{tab::arch}
\begin{tabular}{ r || c | l ||| c | c | c | c }
 \multicolumn{1}{c}{} & \multicolumn{1}{c}{} & \multicolumn{1}{c}{}  & \multicolumn{3}{c}{\emph{\textbf{Data}}}\\\cline{4-6}
\textbf{Description} & \textbf{References} & \textbf{Supported Technologies} & \emph{\textbf{Presence}} & \emph{\textbf{Coordination}} & \emph{\textbf{Computation}}  \\\hline\hline
Mass customization & \cite{BONEV201558} & Assembly Line & localized & centralized & concentrated \\\hline
Manufacturing service composition & \cite{XUE201628} & Assembly Line & localized & centralized & concentrated \\\hline
Computer integrated manufacturing & \cite{DELARAM201896} & Assembly Line & localized & centralized & concentrated \\\hline
Collaborative manufacturing & \cite{doi:10.1080/0951192X.2016.1185155} & Assembly Line & localized & centralized & distributed \\\hline
Dynamic manufacturing reconfiguration & \cite{8295235} & Assembly Line, NCS & localized & centralized & distributed \\\hline
Cloud manufacturing &\makecell[c]{ \cite{doi:10.1080/0951192X.2015.1066861} \\\cite{doi:10.1080/0951192X.2015.1125766}} & Assembly Line, NCS & ubiquitous & \makecell[c]{hierarchical\\centralized} & \makecell[c]{distributed\\concentrated} \\\hline
%Fluid flow modeling & \cite{7152942} & Assembly Line &&&\\\hline
%Ubiquitous manufacturing & \makecell[c]{\cite{doi:10.1080/0951192X.2014.1003411} \\\cite{doi:10.1080/0951192X.2015.1032355}} & Assembly Line, IIoT &&&\\\hline
NCS SW reuse and integration & \cite{CAMPANELLI201547} & NCS & localized & centralized & concentrated \\\hline
Control-based robot navigation & \cite{8241718} & Industrial Robots & localized & centralized & concentrated \\\hline\hline
Deterministic consumer services & \cite{7498096} & IIoT & ubiquitous & centralized & concentrated\\\hline
Green IIoT & \cite{7785890} & IIoT & ubiquitous & centralized & concentrated \\\hline
Service-oriented modeling & \makecell[c]{\cite{SADOK201558}\\\cite{doi:10.1080/0951192X.2017.1392615}\\\cite{doi:10.1515/auto-2015-0017}} & IIoT, WSAN, NCS & ubiquitous &\makecell[c]{hierarchical\\centralized}& distributed\\\hline
%Distributed demand/response & \cite{7750612} & IIoT &&&\\\hline
%Real-time project management& \cite{doi:10.1080/0951192X.2015.1130242} & IIoT &&&\\\hline
%Networking middleware & \cite{SADOK201558} & IIoT, WSAN, NCS & ubiquitous & hierarchical & distributed \\\hline
%Hybrid wireless communications/data & \cite{electronics7120400} & \makecell[c]{IIoT / ICPS,\\M2M Communication} & ubiquitous & hierarchical & distributed \\\hline
Hierarchical data communication & \makecell[c]{\cite{electronics7120400} \\\cite{7885069} \\\cite{8333734}} & \makecell[l]{IIoT / ICPS, WSAN, NCS, \\M2M Communication} & ubiquitous & hierarchical & distributed \\\hline 
Communication harmonization & \cite{7883994} &  \makecell[l]{IIoT / ICPS, \\M2M Communication} & ubiquitous & centralized & concentrated\\\hline
%Process control system migration & \cite{doi:10.1080/0951192X.2017.1392615} & WSAN, NCS & ubiquitous & centralized & distributed\\\hline
Plant-wide process monitoring & \cite{7358111} & WSAN, NCS & ubiquitous & centralized & concentrated \\\hline
Wireless networked control systems & \cite{7466847} & WSAN, NCS & ubiquitous & centralized & concentrated \\\cline{4-6}
\end{tabular}
\end{center}
\end{table*}

The data management information is displayed in Table \ref{tab::arch}. We aim at extracting three specific data properties, in order to understand the recent trends in recent I4.0 architectural design. Meanwhile, we also identify the major supported technological enablers per architectural design. The three data properties we focus are the following:

\begin{enumerate}
\item \emph{Data presence}: Data can be acquired from specifically defined, localized sources, or from pervasive data generators. We label the first category as \emph{localized} data presence. This category usually includes (but is not limited to) data generation sources such as fixed robotic manipulators in a factory environment, stationary network controllers, servers, office workstations, and fieldbus masters. We label the second category as \emph{ubiquitous} data presence. This category includes  (but, again, is not limited to) workers' portable devices, IIoT enablers, sensors and actuators with uncertain communication patterns and online third party data sources (e.g., via Internet). 
\item \emph{Data coordination}: Coordination of the industrial processes, based on the input data, can be performed by global or local process (or network) managers. In the case of involvement of local managers, usually hierarchy is applied, where the coordination is structured among different layers of managers. We label the first case of global managers as \emph{centralized} coordination and the second case of local managers participating in hierarchical managing as \emph{hierarchical} coordination. The most usual trade-off that exists between the different types of coordination is balancing the effect of central control on the network over the minimization of important metrics such as end-to-end data delivery delay and energy consumption.
\item \emph{Data computation}: Computation tasks over the received data can take place either on central entities with significant computational abilities (which may or may not coincide with the coordination managers) or on a large part, or all, of the devices available in the architectural design. We label the first method as \emph{concentrated} computation and the second method as \emph{distributed} computation. Following the concentrated computation model, implies stronger computational power located on single computational components, while following the distributed computation model implies that computation components are located on different networked computers (usually of lower computational ability compared to the concentrated computation case), which communicate and coordinate their actions by passing data to one another. As with typical distributed systems, the three significant characteristics of distributed computation in I4.0 are concurrency of computations, lack of a global clock, and independent failure of the computational devices. For this reason, usually, a failure in the concentrated computation case can lead to much higher failure impact on the industrial processes.
\end{enumerate}

A conclusion drawn by the information extracted by the relevant articles and provided in Table \ref{tab::arch} is that the architectural trends can be classified in two distinct categories, each one with their respective data management philosophy. On the one hand, we have a set of architectures dealing mostly with localized data, coordinating the industrial devices in a centralized manner and providing a mix of either concentrated or distributed computing. The basic data enabling technologies for those architectural designs are the assembly line and the industrial robots. On the other hand, we have a set of architectures dealing mostly with ubiquitous data presence, with a  twist on coordination towards a hierarchical manner, providing again a mix of centralized and distributed computation. The basic data enabling technologies for those architectural designs are IIoT / ICPS, and WSAN. This distinction in two categories of architectural data management makes clear also the diversity of the two research fields (Communications/Networking/Computation and Industrial/Manufacturing/Automation), as well as the necessity of a convergence between the two fields in order to address the I4.0 requirements with common tools and methodologies. This fact is identified as an open challenge for the future and is also presented in section \ref{sec::convergence}.

\subsection{Architectures focusing on assembly line and industrial robots}

In \cite{BONEV201558}, the authors introduce an architecture for the design and customization of product families. Specifically, they design a formal computer-assisted approach that addresses the requirements for the design of product family architecture as identified by academia and industry. The suggested design is based on formal computational models which employ related centralized methods, not leaving much space for ubiquitous data presence and coordination.

In \cite{doi:10.1080/0951192X.2016.1185155}, the authors present an architectural design for interoperable end-to-end manufacturing which guarantees seamless interoperability, thus ensuring proper communication and data exchange between all the partners in a manufacturing network throughout the entire manufacturing life cycle, from supplier search to manufacturing execution and monitoring. In terms of data presence, although the data can lie on different physical locations (e.g., different factories) we consider the layout as localized, since it is perfectly defined beforehands where, when and how the data will be accessed by the platform provided in the architecture. 

\begin{figure}[t!]
\centering
\includegraphics[width=\columnwidth]{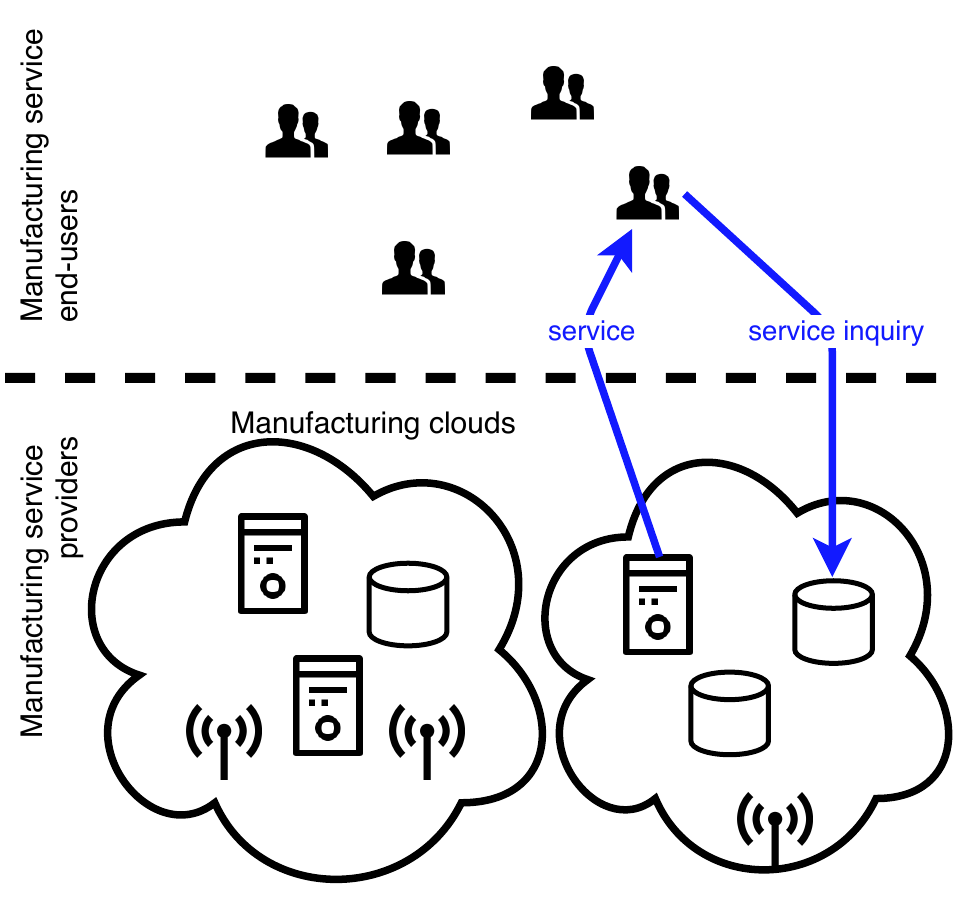}
\caption{Decentralized architecture for cloud manufacturing \cite{doi:10.1080/0951192X.2015.1066861}.}
\label{fig::cloudarch}
\end{figure}

Cloud manufacturing has been a vibrant field for architectural research. In \cite{doi:10.1080/0951192X.2015.1066861}, the authors argue that existing cloud manufacturing models operate in a centralized way through a cloud manufacturing platform, the management of which is identified as a critical part of the manufacturing cloud operation, and strive for decentralization. In fact, they propose a decentralized network architecture which builds upon the concept of autonomous work systems for use as service providers (Fig.~\ref{fig::cloudarch}). In this design, data can be generated from various sources, even from third-party online knowledge clouds and the various computations can happen in different cloud services, with a decentralized coordination, distributively among the users. In \cite{doi:10.1080/0951192X.2015.1125766} the authors introduce the concept of a cloud manufacturing framework with auto-scaling capability, aiming at providing a systematic and rapid development approach for building cloud manufacturing systems. Contrary to \cite{doi:10.1080/0951192X.2015.1066861}, the design of \cite{doi:10.1080/0951192X.2015.1125766} provides a structured and centralized bulletin board data exchange mechanism, serving specifically defined data. However, due to the fact that workers are involved in the design, the number of which varies from time to time (due to the auto-scaling mechanism of the cloud manufacturing framework), the data presence can be considered as ubiquitous also in this case.

In \cite{XUE201628}, the authors investigate how to find the optimal manufacturing service composition path from a service composition network. In order to satisfy the specific demands of manufacturing service composition, they provide a design which solves two problems: how to design the appropriate QoS evaluation model to depict the manufacturing service composition based on networked collaboration, and how to improve the existing service composition method to deal with the rapid increase of candidate service composition solutions. The structure of service supporting system they propose is highly centralized, with regulated coordination and computation of the data resources, which come on the one end from manufacturing, lab and management sources, and on the other end from service requestors.

In \cite{8295235}, the authors introduce a service oriented architectural framework that supports a new programming paradigm for designing dynamic distributed manufacturing systems. The framework supports concurrency and reactivity of multiple computing machines that run data computations asynchronously with each other. Each machine is potentially running concurrent software behaviors that need to execute in synchronously with each other. The entire coordination of the operations is regulated by a master controller. 

In \cite{CAMPANELLI201547}, the authors design an architecture to integrate modules of two industrial standards, IEC 61131-3 and IEC 61499, allowing the exploitation of the benefits of both. The proposed architecture is based on the coexistence of control software of the two standards. As both standards refer to PLCs and control systems, the presence, coordination and computation of data are fundamentally concentrated.

In \cite{DELARAM201896}, the authors propose a layered architecture which covers five critical aspects of computer integrated manufacturing, separated in five architectural layers: physical, functional, managerial, informational and control. Although the holistic design of this architecture is hierarchical and each layer is a separate entity from the other layers, the intra-layer functions regarding coordination and computation can be considered focused on central entities.

In \cite{8241718}, the authors present a general framework for mobile robot navigation in industrial environments in which the open-loop behavior of the robot and the specifications are based on automata. A modular supervisory controller ensures the correct navigation of the robot in the presence of unpredictable obstacles and is obtained by the conjunction of two supervisors: a first one that enforces the robot to follow the path defined by the planner and a second one that imposes other specifications such as prevention of collisions, task and movement management, and distinction between permanent and intermittent obstacles. The data related components are highly centralized both in the planning and in the supervising process of the robot.

%To meet future manufacturing support and sustainability requirements two ubiquitous manufacturing designs have been proposed in \cite{doi:10.1080/0951192X.2014.1003411, doi:10.1080/0951192X.2015.1032355}. The authors present 
\subsection{Architectures focusing on IIoT / ICPS, and WSAN}

\begin{figure}[t!]
\centering
\includegraphics[width=\columnwidth]{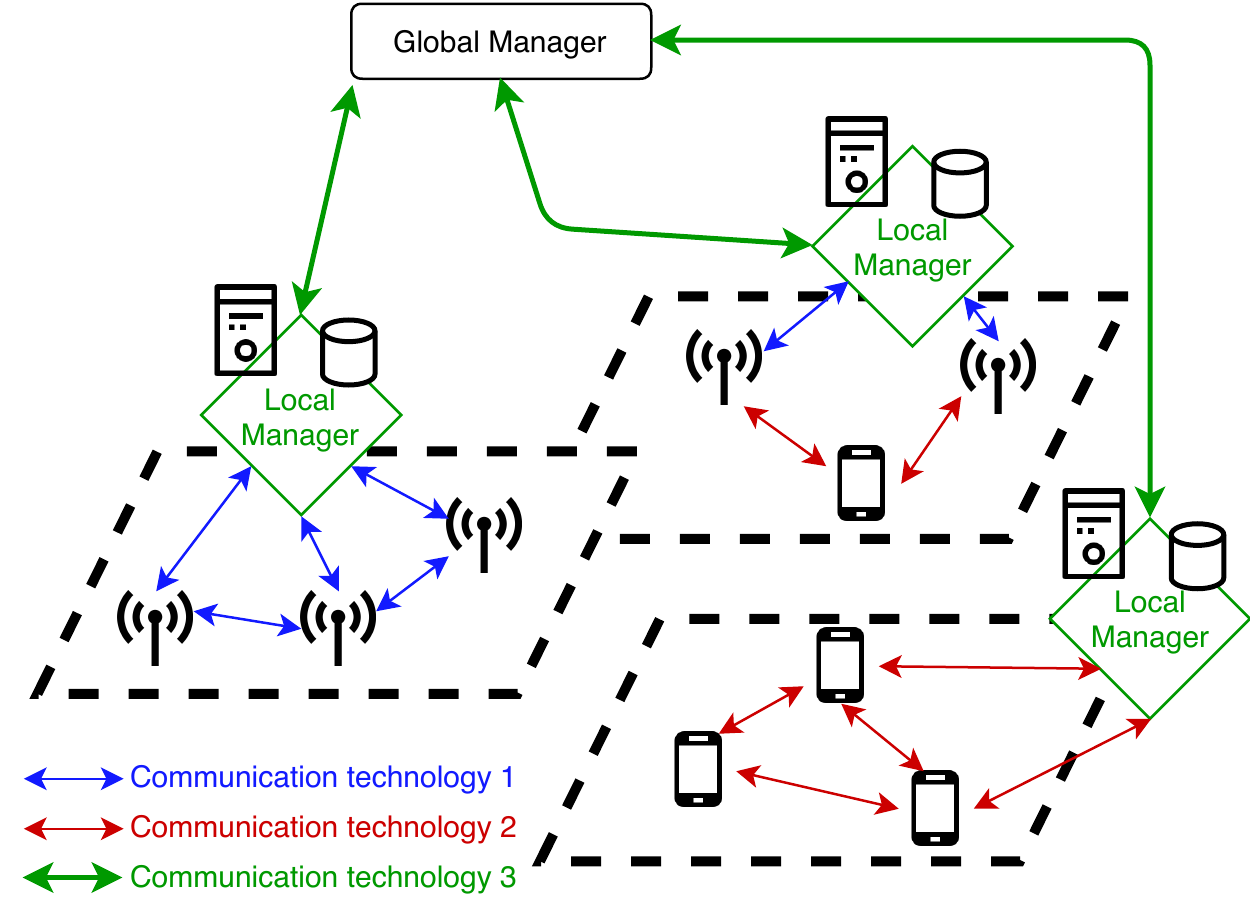}
\caption{Hybrid wireless communication and data management architecture \cite{electronics7120400}.}
\label{fig::mari}
\end{figure}

In \cite{electronics7120400}, the authors introduce a hybrid wireless communication and data management architectural design (Fig.~\ref{fig::mari}). This design is coined as hybrid due to the fact that it is actually a multi-tier network architecture in which distributed communication and data entities interact in order to coordinate their decisions in a hierarchical manner and ensure the correct operation of the whole network. Devices scattered in the network deployment have the ability to perform local computations, lightening the burden of local and global managers by offloading data and computation. The architecture is designed to support ubiquitous data existence in various types of industrial environments.

\begin{figure}[t!]
\centering
\includegraphics[width=\columnwidth]{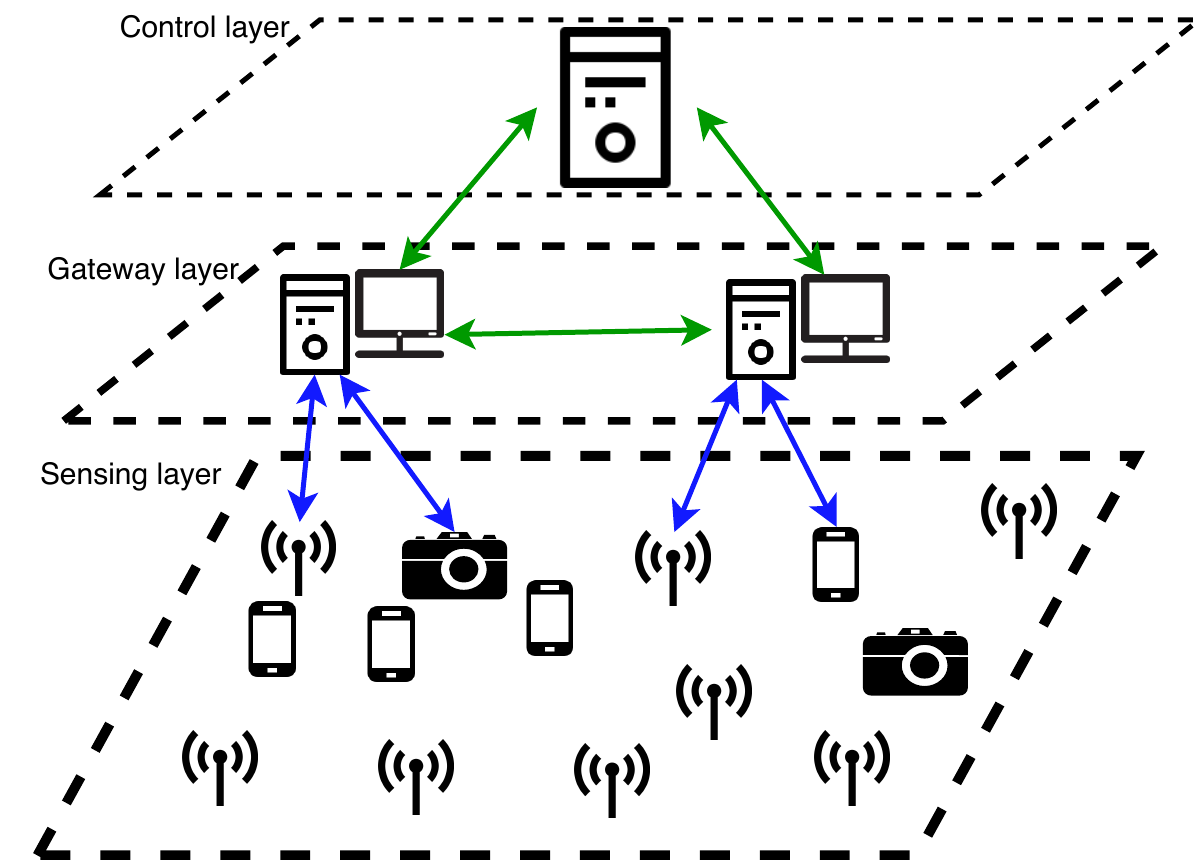}
\caption{Three-layered IIoT architectural design \cite{7785890}.}
\label{fig::greenarch}
\end{figure}

In \cite{7785890}, the authors present an energy-efficient architecture for IIoT deployments, which consists of an IIoT nodes domain, RESTful service hosted networks, a cloud server, and user applications. This architecture focuses on the IIoT domain where large amounts of energy are consumed by large numbers of nodes. The architecture includes three layers: the IIoT layer, the gateway layer, and the control layer (Fig.~\ref{fig::greenarch}). Unlike other hierarchical deployment schemes like \cite{electronics7120400}, in this architecture direct communications between IIoT nodes are not allowed. Also, the gateway nodes are always used as central computation entities and the control node as coordination entity, allowing IIoT nodes to not necessary to implement sophisticated hardware or run complicated routing mechanisms, thus reducing computational complexity and system cost.

In \cite{7498096}, the authors argue that a convergence between deterministic industrial networks and best effort IIoT should occur and support low latency and jitter, and based on this argument, they provide an architectural design for a deterministic IIoT core network consisting of many simple deterministic packet switches configured by an SDN control plane. Although there is a pervasive presence of data due to the IIoT support, the determinism imposes a highly centralized data coordination and schedules computation.

\begin{figure}[b!]
\centering
\includegraphics[width=\columnwidth]{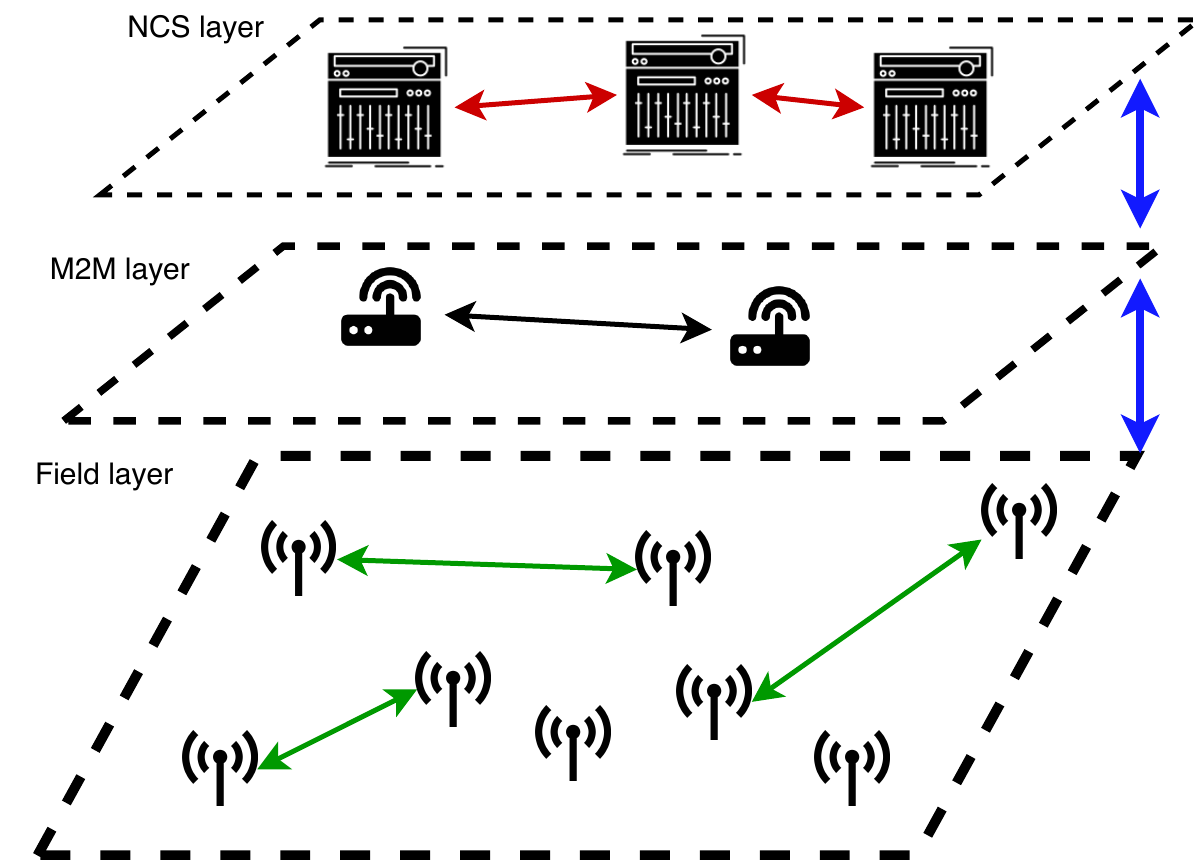}
\caption{Closed loop architectural design for automated wireless NCS \cite{7466847}.}
\label{fig::ncsarch}
\end{figure}

In \cite{7466847}, the authors propose a closed loop design in order to facilitate the deployment of fully automated wireless networked control systems. The topology of the architecture consists of a plant system having sensor and actuator nodes, a controller system having input and output nodes, an intermediate network system having interconnected nodes, and wireless communication links for the information transfer between the different nodes (Fig.~\ref{fig::ncsarch}). The data presence in this setting is ubiquitous, as data can be received by a wide number of sensors placed in the network. However, both the computation and the coordination is taking place centrally at the controller system, which uses the input nodes to receive information and the output nodes to provide controller decisions.

Service-oriented modeling has attracted a lot of attention in the I4.0 architectural design community. In \cite{SADOK201558}, the authors suggest a service oriented architecture which exposes objects' capabilities by means of web services, thus supporting syntactic and semantic interoperability among different technologies. WSAN devices and legacy subsystems cooperate while orchestrated by a manager in charge of enforcing a distributed logic. The architecture supports dynamic spectrum management, distributed control logic, object virtualization, WSANs gateways, a SCADA gateway service, and data fusion transport capability. In order to implement those functionalities, a hierarchical coordination scheme has been followed with different kinds of managers provided as reusable core software components. The middleware's virtualization layer enables the architecture to support pervasive data access and management. In \cite{doi:10.1080/0951192X.2017.1392615}, the authors suggest another service oriented architecture, targeting structured migration of process control systems. The argue that although today's control systems are typically structured in a hierarchical manner, there are nevertheless non-resolved challenges with respect to various fundamental migration functionalities. The suggested approach combines distributed computation abilities with a per-layer centralized coordination, handling data coming from ubiquitous data sources like WSANs. A particular note about this design is that the coordination can also be viewed as decentralized, if we consider the entire system definition and if we do not examine each architectural layer individually. In \cite{doi:10.1515/auto-2015-0017}, the authors argue that the scope of I4.0 shall be defined by considering the major value chains and in order to achieve this they introduce a design and the basic process to achieve a reference model for I4.0 service architectures. The design relies upon the assumption that a reference model should take into account existing reference models for distributed processing as well as those of the Internet of Service and IIoT. This architecture provides a computational modularity which enables distribution through functional decomposition of the system into objects which interact at interfaces.

In \cite{7885069, 8333734}, the authors introduce two different, yet complementary hierarchical data transmission architectural designs for WSAN and smart factories. Those architectures constitute an ideal example of pervasive data generation, as data are received from a wide variety of stationary and mobile sources, such as automatic guided vehicles, mobile workers' devices and WSANs. Hierarchical coordination lies at the core of those designs as well as the decentralized computation through subnetworks formation, leader election algorithms and mobile intelligence units. 

In \cite{7358111}, the authors introduce a distributed modeling framework for plant-wide process monitoring. Based on this framework, the plant-wide monitoring process is decomposed into different blocks, and statistical data models are constructed in those blocks. The data obtained from different blocks are integrated through a centrally located decision fusion algorithm. Due to the large volume of the pervasive plant-wide data generation, the authors note that unlike traditional industrial processes, several new data characteristics should be paid attention to in the plant-wide process: the data volume in the plant-wide process is larger, different types of data can be obtained, sampling rates of process variables are always different from each other, and the density of the collected data from the plant-wide process may be quite low.

Finally, in \cite{7883994}, the authors, rather than presenting a concrete architecture, are providing the future I4.0 architectural insights, based on current designs and future trends, focusing on TSN and 5G designs. Although their analysis includes different vertical integration layers (which enable ubiquitous data presence), it seems that the data coordination and the relevant computations are considered centralized, for the sake of ultra-high reliability.

\section{Data Aspects of I4.0 Technologies and Services} \label{sec::outline}

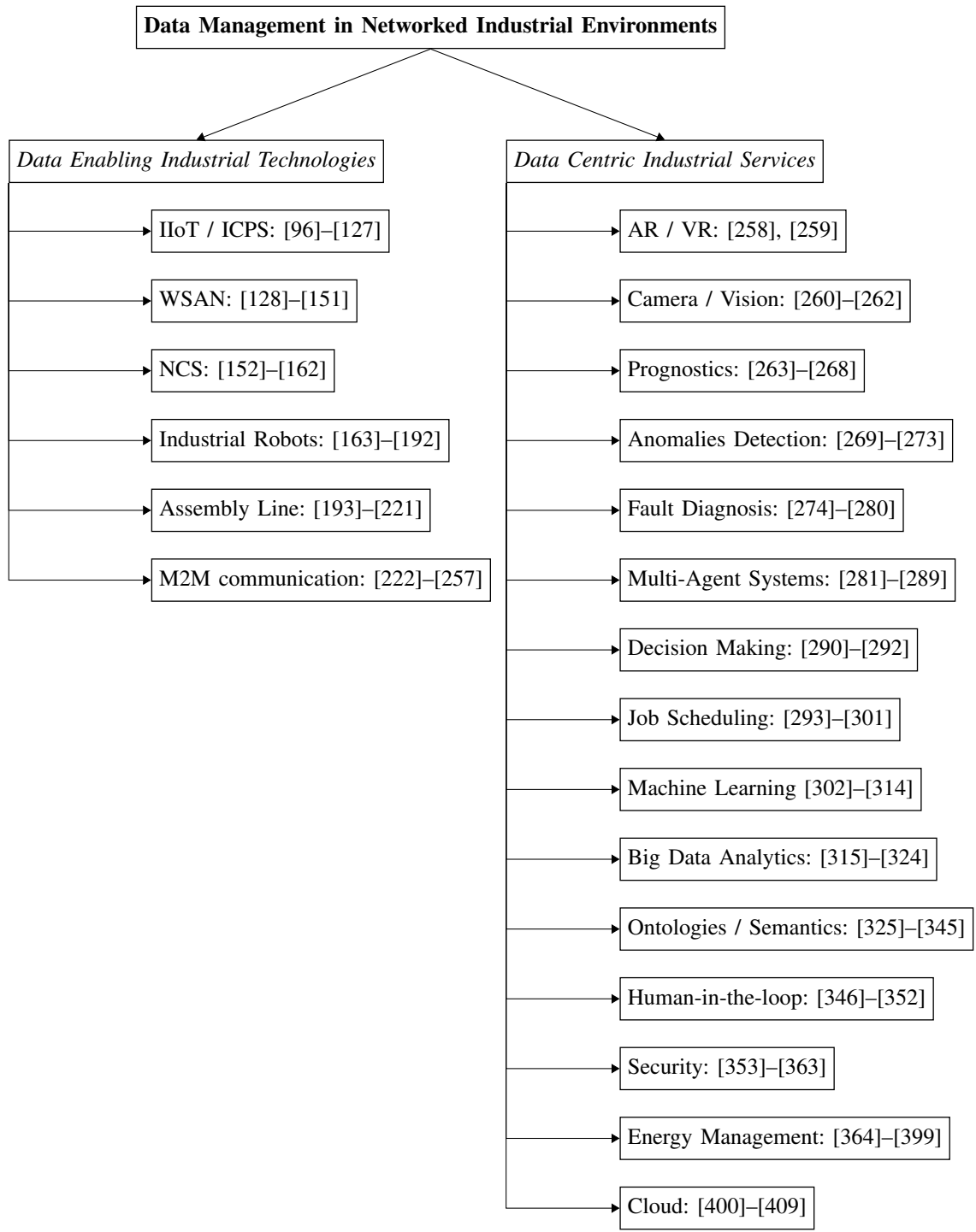
\begin{figure*}[h!]
\centering
\begin{forest}
      for tree={
        line width=0.5pt,
        draw=black,
        fit=rectangle,
        edge={color=black,>={Triangle[]}, ->},
        if level=0{%
          l sep+=1cm,
          for descendants={%
            calign=first,
          },
          align=center,
          parent anchor=south,
        }{%
          if level=1{%
            parent anchor=south west,
            child anchor=north,
            tier=three ways,
            align=center,
            for descendants={%
              child anchor=west,
              parent anchor=west,
              align=left,
              anchor=west,
              xshift=-20pt,
              edge path={
                \noexpand\path[\forestoption{edge}]
                (!to tier=three ways.parent anchor) |-
                (.child anchor)\forestoption{edge label};
              },
            },
          }{}%
        },
      }
      [\textbf{Data Management in Networked Industrial Environments}
        [\emph{Data Enabling Industrial Technologies}
          [IIoT / ICPS: \cite{ZHAI20168, 7486049, 7442163, 7422078, CIVERCHIA20174, 7801816, Santillan2017, 7990138, 7882646, 7878578, 7926931, 8272467, 8291111, 8267270, 8270627, 8186237, 8187647, 8241857, 8362975, 8281615, 8260892, doi:10.1080/0951192X.2017.1407445, THRAMBOULIDIS201592, doi:10.1515/auto-2014-1153, doi:10.1515/auto-2015-0066, doi:10.1515/auto-2016-0065, 8025415, 7955049, 8241845, 8055624, 8412519, Raptis2018}\\ 
            [WSAN: \cite{7039243, 7018937, 7297795, ESKOLA201619, 7359181, 7485858, 7460250, 7576704, 7563316, 7755835, 7888983, 7742966, 7817794, MONTERO201741, 7447766, 7506102, GOMES2017116, 7299629, 8283788, 8291110, 8070398, CAO2018225, 7091856, 7342881}
              [NCS: \cite{7295628, 7410077, 7526316, 7565558, 7723885, 7738402, 7137681, 7973031, 8264779, 8301579, 7275165}
                [Industrial Robots: \cite{6824186, 6936298, 6862044, 7047860, 7298437, 6803999, 7508948, 7289430, 7448900, 7208873, 7470625, 7471445, 8302510, 7934011, 7560600, 7779125, 7792585, 7831386, 7835650, 7864333, 7807322, 7574390, 7569090, 7564480, 8113497, 8170258, 8245858, 8186211, 8082558, 7938664}
                %7812813
                  [Assembly Line: \cite{6979240, doi:10.1080/0951192X.2014.880804, 6871319, 7058453, BRUUN201597, WANG201543, doi:10.1515/auto-2015-0014, doi:10.1515/auto-2014-1157, 7039248, doi:10.1080/0951192X.2014.880811, Diao2016, PAN201667, POORKIANY2016522, 7121030, 7336570, 7529135, POURHASSAN201749, 7790914, doi:10.1080/0951192X.2016.1247989, Liu2017, PENAS201752, 8010300, ELTAIEF2018134, SIERLA201834, RiveraTorres2018, 8283835, CHEN201875, doi:10.1515/auto-2017-0126, KHAJAVI2018100}
                    [M2M communication: \cite{6936873, CHANG2015165, 6880799, 7097039, 6883138, 7517394, 7337450, 7465756, 7345586, 7407396, doi:10.1515/auto-2016-0072, 7362187, 7588154, 7589995, 7961259, BENYAALA2017112, 7883853, 8048552, 7790808, 7790902, 7879239, 7565478, VALLATI20171, SEPULCRE2016121, 7573005, 7546834, 7558198, 8126840, VOGLI201865, 8060997, 8113579, 8242672, 7995068, 7393813, 7781605, 8403102}
                    %7299309
                    %TSCH: CHANG2015165, 7362187, BENYAALA2017112, VALLATI20171, VOGLI201865
                    ]
                  ]
                ]
              ]
            ]
          ]
        ]
                [\emph{Data Centric Industrial Services}
          [AR / VR: \cite{doi:10.1080/0951192X.2013.874589, TATIC20171}
            [Camera / Vision: \cite{6823739, 7131526, 7434623} 
              [Prognostics: \cite{6824783, 7373606, 7412705, DI20181, 7997769, 7520653}
                [Anomalies Detection: \cite{6882174, doi:10.1080/0951192X.2014.961961, doi:10.1515/auto-2015-0060, 7458838, 8106743}
                  [Fault Diagnosis: \cite{7118693, 6915850, 7042841, 7390039, 7933974, 7933998, 8231190}
                    [Multi-Agent Systems: \cite{7104129, CUPEK2016245, doi:10.1080/0951192X.2015.1066862, 7152913, 7593295, 7872487, TERAN201711, 7563318, doi:10.1080/0951192X.2017.1379097}
                    %7778995, 7976383
                      [Decision Making: \cite{KAUR2015151, 7446300, WANG2016158} 
                        [Job Scheduling: \cite{doi:10.1080/0951192X.2014.961548, doi:10.1080/0951192X.2014.964322, doi:10.1080/0951192X.2014.928747, 7103301, doi:10.1080/0951192X.2015.1068450, Hu2017, 7929378, Ni2017, 7842622}
                          [Machine Learning \cite{LUO201681, 7378488, 7572996, 7913594, 8006248, 8301555, 8002611, 8003415, 8241382, 8024058, 8089430, 7999287, 8314728}
                          %8239648
                           [Big Data Analytics: \cite{7111303, doi:10.1515/auto-2016-0022, 7857790, 7835187, 8252775, 8047971, 8291112, FLATH201816, 8231148, Tao2018}
                             [Ontologies / Semantics: \cite{6873317, 7295624, doi:10.1080/0951192X.2014.880809, doi:10.1080/0951192X.2013.874593, PUTTONEN20151041,PEREZGALLARDO201640, doi:10.1515/auto-2015-0094, PALMER201648, THRAMBOULIDIS2016259, 7390286, 7968321, 7762918, 7523919, 7917368, 7569078, HAN201854, 8290684, 8267108, LI201810, 8091285, 8263185}
                             %7911227, 8015195, CHIARELLO2018244
                              % [Knowledge Discovery: \cite{doi:10.1080/0951192X.2014.941403, FRANCALANZA201739, 8295246, 8370886}
                                 [Human-in-the-loop: \cite{LIN2016113, 7733160, doi:10.1080/0951192X.2016.1187295, 7576696, 7837601, 8241351, 7790844}
                                   [Security: \cite{7090973, 7029608, 7506334, 7866869, 8093698, 8270567, URQUHART2018450, LIN201842, 7927473, 8291109, 8283561}
                                     [Energy Management: \cite{6784140, doi:10.1080/0951192X.2014.914631, 6787044, 7097020, 6922531, HAN201619, 7283537, 7588229, 7452389, 7208869, 7588224, 7588226, 7560611, 7588228, 7463484, 7403920, 7588232, 7369957, 7137687, doi:10.1080/0951192X.2016.1185154, 7738397, 7517231, 7898479, 7938652, 7140840, 7539665, doi:10.1080/0951192X.2017.1322220, doi:10.1080/0951192X.2017.1407875, 8241345, doi:10.1080/0951192X.2017.1285429, 8338161, 8272309, 8115313, 8085176, HAN2018205, doi:10.1080/0951192X.2017.1305508}
                                       [Cloud: \cite{7335620, LI2017133, doi:10.1080/0951192X.2017.1314015, doi:10.1080/0951192X.2015.1067912, doi:10.1080/0951192X.2015.1067916, 7415937, 8252786, 8110714, doi:10.1080/0951192X.2017.1407446, LIU201472}
                                       %6908023
                                       ]
                                     ]
                                   ]
                                 ]
                               %]
                             ]
                           ]
                          ]
                        ]
                      ]
                    ]
                  ]
                ]
              ]
            ]
          ]
        ]
      ]
    \end{forest}
    \caption{Taxonomy of I4.0 data management enablers.}
    \label{fig::works}
    \end{figure*}

In this section, we provide a holistic outline of the latest I4.0 data enabling technologies and data-centric services, that were identified through the exhaustive state of the art research, spanning all the way from the field level deep in the physical deployments up to the cloud level. Fig.~\ref{fig::bigpic} visually displays the partitioning of the networked industrial environment building blocks in two fundamental planes: data enabling industrial technologies and data centric industrial services. It is visible that each building block can have thematic and functional overlaps with other building blocks that lie in its proximity. This is natural, and is due to the interplay between modern technologies and services. %In the following we explain which are the main building blocks that we later use for extracting useful data management properties. A more detailed analysis of the reported technological enablers and services, with respect to their data management characteristics, can be found in section \ref{}.
The articles that we have identified and present in this article on I4.0 technologies and services are displayed in Fig.~\ref{fig::works}. In fact, the information presented in Fig.~\ref{fig::works} provides a concise classification in the two categories of the recent research works.

\subsection{Data enabling industrial technologies} \label{sec::techs}

\subsubsection{IIoT / ICPS}

Industrial networked environments are composed of the physical part, which performs the physical processes, and networks of IIoT devices, which perform the computational processes required to control the physical ones. The cyber part of the system is constituted by computational processes, which receive data from the physical processes, calculate the required outputs and apply them to the physical plant \cite{THRAMBOULIDIS201592}, providing and using, at the same time, data accessing and data processing services available on the Internet \cite{doi:10.1515/auto-2015-0066}. Due to the fact that production scheduling is optimized using objective functions based on punctuality criteria such as earliness and tardiness \cite{doi:10.1080/0951192X.2017.1407445}, significant part of those computations are taking place at the edge of the IIoT deployments, transforming edge computing in a fundamental type of computation, with contributions ranging from adaptive transmission optimization \cite{8267270} to multiple gateway optimization \cite{8270627}. Additionally, different IIoT deployments usually incorporate different communication and networking alternatives, such as WIrelessHART \cite{7878578}, RPL \cite{8412519} and 6TiSCH \cite{7926931}, as well as frequent protocol conversions \cite{7990138}, operations which have to seamlessly exchange data with each other. Consequently IIoT and ICPS technologies enable intelligent, adaptive control with seamless vertical, horizontal and dynamic data exchange between heterogeneous platforms and networks, through an exhaustive use of data exchange, coordination and collaboration \cite{doi:10.1515/auto-2014-1153}, as well as through recently proposed techniques like network slicing \cite{8362975}. Important ICPS operations include fault management \cite{doi:10.1515/auto-2016-0065}, clustering analytics \cite{8025415}, reusable software \cite{7955049}, as well as reactive test case generation \cite{8241845} and modular reconfiguration \cite{8055624}. Typical IIoT applications include predictive maintenance \cite{CIVERCHIA20174}, where a successful network configuration is able to determine the condition of the in-service equipment in order to estimate when maintenance should be performed, real-time RFID monitoring \cite{ZHAI20168}, for tracking products in the assembly line. Other research issues include IIoT topology optimization \cite{7486049}, packet scheduling \cite{Santillan2017}, and IIoT network construction and operation under massive multiple-input multiple-output M2M communication \cite{8241857}.

There have been some interesting recent data related advancements in the IIoT domain. In \cite{7442163}, the authors identify the need for data access control along the supply chain, especially when it comes to product data related to sensitive business issues, and they design a scalable industry data access control system that addresses these limitations. In \cite{7801816}, the authors present an industrial data exchange mechanism based on ZeroMQ for the ubiquitous data access in rich sensing pervasive industrial applications. This investigation highlights the major concerns in building a distributed industrial data system in a systematic manner. In \cite{7882646}, identify that most of the current data clustering techniques that could only deal with static data become infeasible to cluster the significant volume of data in the dynamic industrial applications, and introduce an incremental clustering algorithm by fast finding and searching of density peaks based on k-mediods, as a way to find the underlying pattern structures embedded in unlabeled data. Driven by the pursuit of green communication, the authors of \cite{8260892} present a space reserved cooperative data caching scheme for IIoT, where the cache space in a base station is divided into two parts, one is used to store the prefetched data from the servers ahead of the device request time and the other is reserved to store the temporarily buffering data in the wireless transmission queue at the device request time. Timely data delivery is also another crucial data management issue in IIoT, and has been frequently combined with the optimization of other important metrics. For example, in \cite{8187647}, the authors provide a loss tolerant data delivery scheme with low energy consumption and end-to-end guarantees. In \cite{Raptis2018}, the authors present a method for identifying and selecting a limited set of proxies in the IIoT network where data needed by the consumer nodes can be cached, so as to guarantee timely data access. In \cite{8281615} they combine it with MAC layer improvements, in \cite{8186237} with incremental time-triggered data flows, and in \cite{7422078} with a fusion of relaying and data aggregation at the source nodes. Regarding this, there are multiple open challenges to address, such as security concerns (the specific case of DDoS mitigation was addressed in \cite{8291111}), and estimation accuracy \cite{8272467}.

\subsubsection{WSAN} 

WSAN are defined as a group of spatially dispersed and dedicated sensors and actuators for indoor \cite{7576704} and outdoor \cite{ESKOLA201619} monitoring and recording of the physical conditions of the industrial environment. WSAN cooperatively deliver the collected data at a central location via single-hop or multi-hop communication \cite{7091856}. WSANs measure environmental conditions like temperature, sound, pollution levels, humidity, and so on. In fact, WSANs are the base to establish a supervisory control and data acquisition system with the benefits of extending the network boundaries and enhancing the network scalability of the industrial environments \cite{7342881}. Recent research interest in the data-driven industrial WSAN literature has been focused on a number of emerging problems. Localization achieved by using the available plant data in WSAN-enabled industrial environments is one of the problems addressed, both in terms of finding the optimal placement sensor locations in the industrial space space (with Delaunay triangulations \cite{7018937} or particle swarm optimizations \cite{CAO2018225}) and of managing to effectively localize mobile robots \cite{7447766}. The industrial environment that the WSANs operate in is very challenging because of dust, heat, water, electromagnetic interference, and interference from other wireless devices, which make it difficult for current WSANs to guarantee reliable real-time communication. For this reason several communication oriented performance improvements have been achieved. Such improvements include reliable communication slot assignment \cite{7039243}, autonomous channel switching for spectrum sharing \cite{7297795}, synchronization for nodes with imprecise timers \cite{7888983}, and real-time link quality estimation \cite{GOMES2017116}. Cooperative data relaying schemes also facilitate secure and interference-free data management, with recent approaches employing fountain-coding aided transmissions \cite{7359181} and belief function based cooperation \cite{7460250}. Other interesting identified data-driven problems for industrial WSANs include neighbor discovery with mobile nodes based on distributed topology data \cite{MONTERO201741}, network isolation avoidance based on local energy data \cite{7299629}, distributed node clustering based on (among others) node similarity data \cite{7742966}, and coverage data hole healing \cite{8070398}. Data routing improvements are also traditionally a core research aspect, recently with approaches targeting network stability based on nodal data \cite{7506102}, and reliable, SNR-assured, anti-jamming data transfers \cite{8291110}. Cross-layer optimization frameworks have also been proposed for this technological enabler, with SchedEx-GA \cite{7485858} (spanning MAC layer and network layer) attempting to identify a network configuration that fulfills all application-specific process requirements over a topology, and CLOC \cite{7755835}, attempting at maximizing the minimum resource redundancy of the network under system stability and schedulability constraints. Last but not least, data-driven learning with sensing data \cite{7563316}, delay and energy improvements with empirical data \cite{7817794, 8283788} have also emerged as important research directions, especially with the introduction of local clouds in the production process. 

\subsubsection{NCS} 

NCS are control systems wherein the control loops are closed through a communication network. An NCS uses a network as a communication medium to connect the plant to a central controller \cite{7410077}. The defining feature of an NCS is that control data and feedback data are exchanged among the system's components in the form of data through a network. The most important feature of NCS is that they connect cyberspace to physical space enabling the execution of several tasks from long distance. In addition, networked control systems eliminate unnecessary wiring reducing the complexity and the overall cost in designing and implementing the control systems. They can also be easily modified or upgraded by adding sensors, actuators. Usual types of such network communication are fieldbuses like CAN and LON, wired connections like IP/Ethernet, etc. Automated or semiautomated verification of access control is a necessary building block in NCS \cite{7295628}, and sampled-data control has been proven to guarantee their synchronization by reducing the updating frequency of the controller and the network communication burden \cite{8301579}. Due to the difficulty in observing the full relationship among numerous NCS components, high-dimensional and sparsematrices describing partial relationships among them have been recently introduced \cite{7973031}. NCS can also be used to connect different plants with solutions provided to achieve given specifications when there are communication delays and losses in communication networks linking central network controllers and the plants \cite{7526316}. Data-driven network control is known to be one of the most efficient control schemes for complex industrial processes due to the difficulty in obtaining accurate mathematical control models \cite{7565558} and to the frequent existence of nonlinearities and stochastic disturbances \cite{7723885}. In fact, data delivery latency is among the most active topics in the NCS field recently. Networked degradations such as data delivery delay and data dropout can nevertheless cause NCS to fail to satisfy performance requirements, and eventually affect the overall reliability \cite{7137681}. In order to address this problem, NCS can be specified in the form of function blocks through relevant standards such as the IEC 61499 standard, the end-to-end data delivery latency over switched Ethernet of which can be assessed with low complexity techniques \cite{7738402}. Also, delay compensation schemes for NCS using CAN bus \cite{8264779}, as well as energy efficient sampling methods \cite{7275165} have been presented. 

\subsubsection{Industrial Robots}

\begin{figure}[t!]
\centering
\begin{forest}
      for tree={
        line width=0.5pt,
        draw=black,
        fit=rectangle,
        edge={color=black,>={Triangle[]}, ->},
        if level=0{%
          l sep+=1cm,
          for descendants={%
            calign=first,
          },
          align=center,
          parent anchor=south,
        }{%
          if level=1{%
            parent anchor=south west,
            child anchor=north,
            tier=three ways,
            align=center,
            for descendants={%
              child anchor=west,
              parent anchor=west,
              align=left,
              anchor=west,
              xshift=-20pt,
              edge path={
                \noexpand\path[\forestoption{edge}]
                (!to tier=three ways.parent anchor) |-
                (.child anchor)\forestoption{edge label};
              },
            },
          }{}%
        },
      }
      [Industrial Robots
      [Stationary Robots
      [Tracking control \\and correction:\\\cite{7471445, 8302510, 6936298} \\\cite{7298437, 7831386, 7864333} \\\cite{7574390, 7564480, 8186211} \\\cite{8082558}
      [Collaboration: \\\cite{7792585, 8113497}
      ]
      ]
      ]
      [Mobile Robots
      [Localization:\\\cite{7508948, 6862044, 7208873}
      [Navigation:\\\cite{7047860, 6803999, 7448900} \\\cite{7470625, 7835650, 8170258}
      [Collaboration: \\\cite{7560600, 7934011, 7779125, 7569090}
      [Other operations:\\\cite{6824186, 7807322, 7938664}
      ]
      ]
      ]
      ]
      ]
      ]
          \end{forest}
    \caption{Industrial robots: An I4.0 data enabling technology.}
    \label{fig::robots}
    \end{figure}
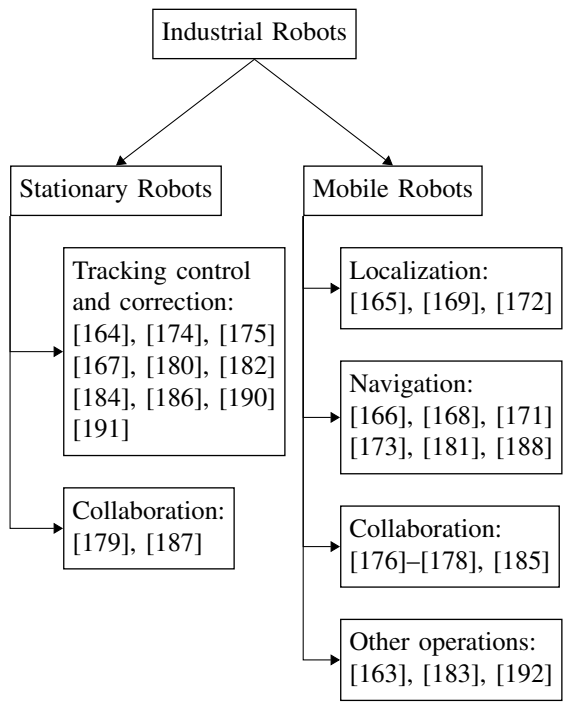

Robot systems have been widely used in industry and also play an important role in human social life \cite{8245858}. Industrial robot research can be classified in two categories (Fig.~\ref{fig::robots}): stationary robots and mobile robots.

Tracking control of robot manipulators is a fundamental and significant problem in robotic industry \cite{8082558}. Tracking control of robotic manipulators with uncertain kinematics and dynamics (gravitational torque, friction torque, moment of inertia and disturbance) is addressed using data-driven observer-based control designs \cite{7471445}, some of which providing convergence of tracking errors \cite{8302510}. Preplanned path tracking corrections of robotic \cite{6936298} or teleoperated manipulators \cite{7298437} can be achieved through iterative learning control algorithms. Smaller robotic parts of larger potential constructs can be controlled distributively through redundant actuation (an example is provided in \cite{7289430}, for a tracking control of a joint). Energy and power efficient methods have also been presented, for a number of cost functions \cite{7831386}. Manipulability optimization of redundant manipulators is shown to be achieved through dynamic neural networks  \cite{7864333}. Neural control is also applied in the case of bimanual robots (which are able to perform more complicated tasks that a single manipulator), resulting in guaranteed stability and precision \cite{7574390}, or in reduced vibrations \cite{7564480}. Data delivery delay is also an important aspect, subject to minimization, shown to be decreased with practical and adaptive time-delay control schemes \cite{8186211}. Coordination and cooperation control for networked mobile manipulators over a jointly connected topology with time delays is another topic that needs fast data delivery in the network \cite{7792585}. Modular design has been proven helpful in the configuration of multirobot cooperation (for example in \cite{8113497} for sewing personalized stent grafts).

Localization of mobile robots in industrial environments is a classic topic that will remain challenging in the I4.0 era. Mobile robots operating in indoor environments \cite{7508948} can be localized with a combination of data coming from heterogeneous sensors, and those operating in outdoor environments \cite{7208873},  with a combination of ambient data (movement dynamics, velocity data, RSSI) High-precision probabilistic localization of mobile robotic fish can be achieved using visual and inertial cues \cite{6862044}. Robot navigation in space is another major topic for data-driven research. Online navigation of humanoid robots has been proven feasible through multi-objective evolutionary approaches \cite{7047860}. Wall-following trajectory control of hexapod robots can be realized via data-driven fuzzy control learned through differential evolution\cite{6803999} and relevant uncertainties can be addressed with decentralizing this control with dynamic controllers \cite{7448900}. Homing (mobile robot returns back to a reference home position) using just the visual information can be implemented by extracting coarse location data with respect to the reference position using a bit encoding algorithms \cite{7470625}. Autonomous exploration using mobile climbing robot allow dangerous tasks to be completed more quickly and more safely than is possible with human inspectors \cite{7835650}. Wireless charging helps mobile robot to become more and more autonomous and navigate easier \cite{8170258}. Except for navigation, several approaches regarding other robot properties have been presented, such as balancing and velocity control \cite{6824186} with in-wheel motors, human behavior transfer to robots through learning by imitation/demonstration \cite{7807322} and visual servo regulation with simultaneous depth identification \cite{7938664}. Robot collaboration and data sharing is also an emerging interesting research issue. Teleoperation control frameworks for multiple coordinated mobile robots through have been proposed using a brain-machine interface \cite{7560600}. A particularly interesting topic in the mobile robots collaboration field field is the collaborative and adaptive data sharing. Collaborative robots are multirobot systems working together for the same industrial task such as robotic assembling. To achieve an efficient collaboration, robots require not only locally sensing the environmental data but also immediately sharing these data with neighbors. However, there exists a dilemma between the large amount of sensory data and the limited wireless bandwidth. The relevant problem of throughput maximization of sensory data sharing in collaborative robots has been studied in \cite{7934011}. Another interesting topic which again necessitates distributed data exchange is the consensus problem. The consensus problem has experienced a fair amount of research interest, aiming at forcing a group of mobile robots to reach an agreement on a quantity of interest such as the rendezvous position, velocity, and heading direction \cite{7779125}. Multiple robots can also collaboratively achieve a common
coverage goal efficiently, which can improve work capacity, share coverage tasks, and reduce completion time \cite{7569090}.
\subsubsection{Assembly Line} 

The assembly process is composed of several data intensive stages, namely, resource identification, resource recognition, data collection, data transmission, data mining, and
feedback control \cite{Liu2017}. Flexibility is critical for manufacturing firms to respond to demand uncertainty and achieve product customization. For example, in automotive plants, vehicles with multiple styles, models, and options can be made on the same production line. Similarly, computers with different configurations are assembled on the same line as well \cite{7121030}. Similar observations are found in many other manufacturing systems, such as appliances, electronics, furniture, food, and are usually described by model-based processes \cite{doi:10.1515/auto-2015-0014}. However, replacing a resource or introducing a new product variant often requires manual integration work and considerable downtime. For this reason, automated systems for manufacturing need to adapt increasingly fast to the new \cite{doi:10.1515/auto-2014-1157}. Data is already playing a crucial role in customized manufacturing, as advanced systems are needed that analyze the assembly and use the plethora of data available at the shopfloor to generate highly flexible assembly sequences. 

In order to increase the requested flexibility and boost the data availability in the production process, assembly lines are being evolved and are featuring new technological improvements. Some fundamental data-enabling advancements for the modern assembly lined include: Sensor data acquisition systems producing large amounts of small volume data \cite{7790914}, (3D) CAD/CAM systems and models producing considerable amounts of large volume data \cite{doi:10.1080/0951192X.2014.880811}, simulation-based systems \cite{doi:10.1515/auto-2017-0126} for rearranging manufacturing facilities targeting material handling and costs minimization producing complex mathematical data \cite{POURHASSAN201749}, digital twins of physical products producing assembly orchestration data \cite{SIERLA201834}, as well as integrated ICPS producing coupled cyber-physical data \cite{PENAS201752}. In \cite{7058453}, the authors introduce a knowledge-based approach exploiting distributed declarative data and cloud computing and target data and software exchange and reuse, maximizing the potential to facilitate new business models for industrial solutions.

Real-time data operations for flexible manufacturing are becoming increasingly popular, are now in the core of the production process and are using different kinds of data. Real-time performance assessment of manufacturing systems by monitoring continuous and discrete variables of different machines is based on data extracted from factory machines \cite{8283835}. Real-time monitoring of the production process is based on data (features) extraction and selection (for example, high-power disk laser welding in \cite{CHEN201875}, with fifteen features extracted). Real-time production exception diagnosis is based on sensor data streams \cite{7336570}. Real-time geometrical re-definitions of products in the assembly line are based on 3D data from CAD systems and models \cite{ELTAIEF2018134}. The same holds for real-time capturing, structuring and assessing the design rationale of product design \cite{POORKIANY2016522}. Real-time coded aperture techniques targeting the alignment process for industrial machinery producing high resolution image data \cite{6871319}.

Some specialized recent contribution on assembly line improvements include the following. In \cite{doi:10.1080/0951192X.2014.880804}, the authors argue that the diversity and uncertainty of data over the dimension, damage degree and remaining life characteristics of used parts make the remanufacturing process route decision more complicated, and they propose a model for finding the optimal remanufacturing route. Due to similar uncertainties of complex mechanical products, the authors of \cite{WANG201543} suggest an assembly quality adaptive control system, in order to improve the products' assembly precision, stability and efficiency. In \cite{BRUUN201597}, the authors adopt a visual product architecture representation in combination with a PLM system data to support the development of a family of products. In \cite{6979240}, the authors introduce an efficient automation and control for a particular type of industry, the conventional cable manufacturing industry, a conventional stranding plant of which takes up approximately 300-400 m$^2$ of space. Last but not least, taking into account that the practice of kitting (to supply the required parts for a single assembly in pre-set containers) provides an alternative to the currently dominant practice of continuous supply line-stocking, the authors of \cite{KHAJAVI2018100} analyze the value of model-based kitting for additive manufacturing.

Several theoretical frameworks have also been proposed. Industrial machines using probabilistic Boolean networks enable the study of the relationship between machine components, their reliability and function \cite{RiveraTorres2018}. Manufacturing systems with batches and duplications can be effectively modeled by timed event graphs and then studied using algebraic tools \cite{8010300}. Time-varying properties of industrial processes can also be seen as data-driven, autoregressive models and be estimated with relevant recursive algorithms \cite{7529135}. Improvements of key features of product manufacturing can be realized via weighted-coupled network-based quality control methods \cite{Diao2016}. Petri nets modeling can augment the performance of event driven systems like intelligent part dispatching using temporal data \cite{7039248}. Integrated process planning and system configuration for machining on rotary transfer machines can be effectively realized through the employment of sophisticated optimization tools \cite{doi:10.1080/0951192X.2016.1247989}. Finally, automatic adaptation of assembly models can be modeled with attributed kinematic graphs \cite{PAN201667}.

\subsubsection{M2M communication} 

Industrial M2M communication refer to direct communication between industrial networked devices using any communications channel, including wired and wireless. Emerging
smart factories are envisioned to be seamlessly integrated with diverse communication technologies. Consequently, production, networking, and communication will become tightly integrated. Cooperation among different sites of a factory or even different factories will be easily possible \cite{doi:10.1515/auto-2016-0072}. The research emphasis on this technological enabler is put less on the large scale network optimization aspects (which are investigated in the rest of the technological enablers) and more on the device to device communication links, channels, transmissions and one hop data exchanges. The exact contributions range from the lower technological level of circuit network model design \cite{6880799}, up to the higher technological levels of antenna design \cite{7565478}, filtering \cite{7546834}, multiplexing \cite{8126840}, interference management \cite{7589995} and others. Particular attention has been paid on guaranteeing the QoS of the subsequent data delivery over the communication media, through various methods, such as function splitting between delay-constrained data delivery and resource allocation \cite{7517394}, redundant communication schemes \cite{SEPULCRE2016121}, or precise communication and network modeling \cite{8048552}. Optical communications have also started penetrating the industrial sector, especially for moderate and high data rates with enhanced security (due to the spatial confinement of optical links) for both short \cite{7097039} and longer ranges \cite{8113579}, however their full potential remains to be unlocked, as the cost of optical equipment is still high \cite{7995068}.

The M2M Communication configuration has a direct impact on the efficiency of the industrial network data management, and especially on specific sensitive data-related metrics. Those metrics are fundamental operatives of the I4.0 and are guaranteeing the smooth function of resource-intensive industrial applications. Some indicative examples where the impact of communication scheme is highly beneficial are the following: self-triggered sampling schemes for NCS targeting low data losses and delays \cite{7337450}, statistical dependences management in channel gains of industrial WSAN targeting efficient data routing \cite{7465756}, phase-sensitive sensing and communication targeting safety-critical data distribution \cite{7781605}, mmW deployments targeting large number of data hops \cite{7879239}, field-oriented network control decoupling targeting effective machine operation \cite{7883853}, and optimized cooperative multiple access techniques targeting efficient resource sharing \cite{8242672}.

A useful standardized recent data enabling communication mechanism is a recent extension of IEEE 802.15.4. Several studies have highlighted that the IEEE 802.15.4 communication standard presents a number of limitations such as low reliability, unbounded packet delays and no protection against interference, that prevent its adoption in applications with stringent requirements in terms of data reliability and latency \cite{DEGUGLIELMO20161}. For this reason, IEEE has released the 802.15.4e amendment that introduces a number of enhancements to the MAC layer of the original standard in order to overcome such limitations. Following this release, there is a constant flow of research on improving various aspects of the amendment. This part of research includes a great number of works on the M2M communication technological enabler, and more specifically concentrated on three of the main 802.15.4e MAC operation modes, Time Slotted Channel Hopping (TSCH), Deterministic and Synchronous Multi-channel Extension (DSME) and Low Latency Deterministic Network (LLDN) (for more details on the functions of those modes, the reader can consult \cite{DEGUGLIELMO20161}). Regarding the TSCH mode, the main research focus has been recently placed on synchronization, with some techniques using learning and prediction data from neighboring nodes \cite{CHANG2015165}, and other techniques using mutual synchronization of distributed nodes \cite{BENYAALA2017112}, as well as on fast network joining algorithms \cite{VOGLI201865}. Regarding the DSME mode, improved network formation has been studied in \cite{VALLATI20171}. Regarding the LLDN mode, significant efforts have been invested in transforming the standard compatible for ultra-low latency applications, where the critical data need to be delivered with high reliability \cite{7362187}.

Another widely used data enabling technology used for data management in industrial environments is the IEEE 802.11 WLAN and its various amendments. The IEEE 802.11 standard revealed effective since it is able to provide satisfactory performance for several industrial applications in which tight requirements in terms of both timeliness and reliability are encountered \cite{7345586}. Specifically, the possibility of implementing ad hoc data management schemes as well as infrastructure configurations, renders it very convenient. Here the emphasis is put on several important aspects. The first aspect is seamless redundancy to improve reliability through reference architectures \cite{7393813}, experimental campaigns \cite{7790902} and joint interference prevention \cite{8060997}. The second aspect concerns soft real-time control applications where the relevant constraints are met through efficient bandwidth management \cite{7961259}, as well as enhanced communication determinism \cite{7790808}. The third aspect is dynamic rate selection algorithms, where data is delivered within the deadlines, while transmission error is minimized \cite{7588154}.

Other data enabling communication technologies include: CAN with jitterless communication via stuff bits prevention \cite{6936873}, OPC-UA with enhanced throughput increased via RESTful architecting \cite{7407396, 8403102}, EtherCAT with very short cycle times via priority-driven swapping-based scheduling of aperiodic real-time data \cite{6883138}, ISA100.11a with increased reliability via adaptive channel diversity \cite{7558198}, WIrelessHART for harsh industrial environments \cite{7573005}. Table \ref{tab::m2m} displays an overview of selected references regarding specific communication technologies.

\begin{table}[t!]
\begin{center}
\caption{Standardized data enabling communication technologies.}
\label{tab::m2m}
\begin{tabular}{ r | l }\hline
\textbf{\makecell[r]{Technology}} & \textbf{Articles}\\\hline\hline
\makecell[r]{IEEE 802.15.4e}	& \cite{CHANG2015165, BENYAALA2017112, VALLATI20171,VOGLI201865, 7362187}	\\\hline
\makecell[r]{IEEE 802.11(a/n)}	& \cite{7345586, 7393813, 7790902, 8060997, 7961259, 7790808, 7588154}	\\\hline
EtherCAT 		& \cite{6883138}	\\\hline
CAN 		& \cite{6936873}	\\\hline
OPC-UA 		& \cite{7407396, 8403102}	\\\hline
ISA100.11a 		& \cite{7558198}	\\\hline
WirelessHART 		& \cite{7573005}	\\\hline
\end{tabular}
\end{center}
\end{table}

\subsection{Data centric industrial services} \label{sec::services}

\subsubsection{AR / VR} 

There have been very few works on augmented reality (AR) and virtual reality (VR) services. Typically, those services require large volumes of video data which are processed centrally with high computational overhead. In \cite{doi:10.1080/0951192X.2013.874589}, the authors introduce a context-aware augmented reality assisted maintenance system, in which industrial users can add and arrange various contents spatially, e.g., texts, images and CAD models, and specify the logical relationships between the AR contents and the maintenance contexts. The data in this system are stored in a context database of the context management module. A context sensing module acquires raw data from the users and various physical sensors in the environment, and interprets the raw data to obtain low-level contexts. The sensor interpreter obtains and interprets data from the physical sensors. For example, it processes the raw images captured by the cameras, and outputs the marker ID and transformation matrix. The data processing is conducted offline on large volumes of acquired data. In \cite{TATIC20171}, the authors apply AR technologies for the improvement of occupational safety in industrial environments. The application is installed on workers' mobile devices that are used as the input and output of the system. All the necessary data are stored in a central database that is accessed by the application whenever required. The system is personalized according to skills of a worker by taking into account his professional training and work experience. Depending on that it is determined the amount of data to be displayed to a worker helping even less skilled workers to perform a task. Therefore, in this case although the data presence is localized, the data processing is distributed.

\subsubsection{Camera / Vision} 

There have been some works which use camera and vision technologies for efficient pattern recognition, fault estimation and template matching. In \cite{6823739}, the authors develop a data-driven decoupling feedforward control scheme with iterative tuning to meet the challenge of the crosstalk problem in MIMO motion control systems. This scheme is data-driven in the sense that, unlike typical model-based approaches of this field, it uses an iterative tuning which uses the available data to overcome the practical obstacles in obtaining an accurate dynamic model. The authors show that through the beneficial use of data and with only one measurement data collection, the decoupling control scheme can reduce the effect of the crosstalk with a decrease of two orders of magnitude ($10^{-8} \rightarrow 10^{-10}$). In \cite{7131526}, the authors present two estimator designs for WSANs in multi-target tracking under signal transmission faults due to the uncertainties in the surrounding environmental conditions. In \cite{7434623}, the authors describe a model-based template matching system, which is robust to undergo rotation and scaling variations. The data used as input in the system are comprised of image data, and, in fact, the authors test the system with different categories of image data, through three diverse datasets: logos and badges, image patches, and PCB components.

\subsubsection{Prognostics}

Prognostics engineers face various situations regarding collected data from the past, present, or future behavior, and have to come up with efficient data-driven solutions. Generally, the modeling of data-driven prognostics has to go through necessary steps of learning and testing. First, raw data are collected from machinery and are preprocessed to extract useful features to learn degradation behavior. Second, in the test phase, the learned model is used to predict future behavior and
to validate model performance. An example of prognostics operations in industrial environments is systems health management, an enabling discipline that uses
sensors to assess the health of systems, diagnoses anomalous behavior, and predicts the remaining useful performance over the life of the asset \cite{7520653}. In \cite{6824783}, the authors present a new approach for feature extraction based on vibration data, targeting accurate prognostics for machinery health monitoring. The main breakthrough of the paper is the mapping of raw vibration data into monotonic features with early trends, which can be easily predicted. The data collection and processing is concentrated on central computation entities. The contribution is naturally data-driven and the authors strive for a good balance between model accuracy and complexity. Prognostics also present a widespread application in network-based industrial processes, with \cite{7373606}, where combined fault-tolerant and predictive control is introduced and \cite{7997769}, where a weighted linear dynamic system for nonlinear dynamic feature extraction is proposed. In those works, the authors try to identify the considerable redundancy and the strong correlations between data as well as to manage the random noises present at data. Other interesting data-driven industrial prognostics applications include \cite{7412705}, which presents an extended prediction self-adaptive controller employing graphical programming of industrial devices for controlling fast processes, and \cite{DI20181}, which investigates fault prediction of power converters in industrial power conversion systems.

\subsubsection{Anomalies Detection}

Considering the aspect of data management, current anomalies detection approaches are either centralized and complicated or restricted due to strict assumptions, a fact that renders them difficult to apply on practical large scale networked industrial systems. The accommodation of high rates of data capture and total data volume generated by complex WSANs that typically monitor industrial systems pose one of the main challenges for online anomalies detection. The paper \cite{doi:10.1515/auto-2015-0060} outlines such centralized data-driven systems for anomalies detection for ICPS using several use cases from industry. Based on data, these systems extract most necessary knowledge about the diagnosis task. Another ICPS-enabled work is \cite{8106743}, in which the authors present an anomaly detection approach for ICPS based on zone partitioning. Additionally, in \cite{7458838}, an online two-dimensional changepoint detection algorithm for sensor-based anomalies detection is proposed. Interestingly enough, in \cite{6882174}, the authors introduce a distributed general anomaly detection scheme, which uses graph theory and exploits spatiotemporal correlations of physical processes to carry out real-time anomaly detection for large scale networked industrial sensing systems. Finally, in \cite{doi:10.1080/0951192X.2014.961961}, a work of different flavor, the authors display the concept of early problem identification in collaborative engineering with different product data
modeling standards.

\subsubsection{Fault diagnosis}

Fault detection, isolation and reconstruction methods are essential to improve the reliability, safety of the automatic control systems. In \cite{7118693}, the authors develop a model-based fault location method is developed for intermittent connection problems on controller area networks. In this type of networks time critical data are transmitted, hence, the reliability of the network not only has a direct impact on the system performance but also affects the safety of the system operations. In \cite{6915850}, the authors introduce a condition monitoring and fault diagnosis scheme of electric motors for harsh industrial applications. The authors also note that for a real implementation in industry, since the proposed scheme assumes prior knowledge of various data in a motor current spectrum, small additional memory might be required to implement the proposed method. Also sufficient bandwidth of data acquisition is required, particularly for high-frequency signal detection. In \cite{7042841}, the authors discuss some basic properties of the failure rate of redundant reliability systems in industrial electronics applications. They note that the the problem of reliability evaluation of the single components is data related and is not an easy matter, and this is exactly in view of the scarcity of failure data. In \cite{7390039}, the authors design a fault isolation technique based on the $k$-nearest neighbor rule for industrial processes. A notable data related remark on this paper is that the technique focuses on the problem of isolating sensor faults only based on the normal data, without any fault information. In \cite{7933974}, a reconstruction-based method is proposed to monitor nonlinear industrial processes and isolate their fault types. This method includes numerous data operations (such as normal data decomposition and faulty data decomposition), and In the experimental section, monitoring data of an electro-fused magnesia furnace is used to show its effectiveness. In \cite{7933998}, the authors suggest a component analysis algorithm for fault monitoring in industrial processes, and in \cite{8231190} a threshold-free error detection scheme for WSANs. Various data oriented techniques are used by the authors, such as exploitation of the information related to the spatial and temporal relationships among sensor data streams, data correlations and mapping of residual data streams.

\subsubsection{Multi-Agent Systems}

Multi-agent systems have been presented as a suitable service to develop modular, flexible, robust, and adaptive large-scale production lines. However, the classical multi-agent systems are defined by a static hierarchy of data structures, which makes them very difficult to modify \cite{CUPEK2016245}. For example, in \cite{doi:10.1080/0951192X.2015.1066862}, the authors present a software platform structured around a central data repository, containing engineering data and information from ongoing and completed line design projects. The central data repository is used by software agents that allowed the seamless update and use of engineering data. Also, in \cite{7152913}, the authors investigate the tracking control problem of networked multi-agent systems centrally with multiple delays and new characterizations of impulses. Many of the recent works focus on the decentralization of industrial functions and data distribution over a community of distributed, autonomous, and cooperative agents. The application of distributed agent data and services allows the achievement of important features, namely modularity, flexibility, robustness, adaptability, reconfigurability, and responsiveness \cite{7563318}. Some recent ones are the following. In \cite{7104129}, the authors develop a multi-agent system for process and quality control in a laundry washing machines factory. They construct an agentification of the factory's production line and distribute the various types of data among different kinds of agents. In \cite{7593295}, the authors model manufacturing machines as agents, which can collect production data and distributively control the machines. Giving them self-organization capability, machines can be reconfigured for different tasks to achieve the highest resource efficiency. Manufacturing processes are monitored and adjusted by the self-adaptive model when exceptions occur. In \cite{7872487}, the authors propose the modeling and synthesis procedures to obtain optimal decentralized industrial controllers in state-feedback form for distributed agents. \cite{TERAN201711}, presents a multi-agent method for industrial process integration implementing coordination optimization mechanisms that enable distributed agent data exchanges, by using cultural algorithms. In \cite{doi:10.1080/0951192X.2017.1379097}, the authors introduce non-cooperative agents which make decisions based on the capacity allocation and the data of all other agents, thus creating a decentralized feedback loop.

%\cite{7976383}, 
%\cite{7778995}, 

\subsubsection{Decision Making} 

The integration of ubiquitous sensing capabilities of IIoT with the industrial infrastructure of I4.0 can enable the automation of the decision making process inside and outside the shop-floor. The data collected by IIoT systems in smart industries can be used to replace manual employee evaluation systems where there are ample chances of bias. In \cite{7446300}, the authors develop a large-scale data-driven multitask learning and decision-making system, which can quickly coordinate machine actions online for large-scale custom manufacturing tasks. In \cite{WANG2016158}, the authors present a self-organized system with data based feedback, coordination and improved decision making ability. In \cite{KAUR2015151}, the authors propose a model for automated performance evaluation of employees in a smart industry. The model uses the data collected by embedded sensors in smart industrial system to identify various industrial activities of employees. The identified activities are then classified as positive, negative and neutral activities. Here the word ``decision'' refers to the action taken in response to the performance of employees. The proposed model consists of an IIoT network, an information processing system and a central database system. The data collected by the IIoT network are stored in the database and used by the information processing system to infer the useful requested results. Another interesting data enabling entity in this paper is the data conversion block, which is used to classify a particular activity into positive, negative or neutral and to calculate the amount of profit or loss corresponding to positive or negative activity respectively. Finally, a decision making block is automatizing the decision making process using game theoretical tools.

\begin{table*}[t!]
\begin{center}
\caption{Data-driven machine learning services for data enabling technologies.}
\label{tab::ml}
\begin{tabular}{ r | c | c | l}\hline
\textbf{\makecell[r]{Data enabling technology}} & \textbf{Articles} &\textbf{Type of service} &\textbf{Method used} \\\hline\hline
IIoT / ICPS			&\makecell[c]{\cite{LUO201681} \\\cite{8241382}}		& \makecell[c]{missing QoS values prediction \\intelligent IIoT traffic classification}	& \makecell[l]{kernel least mean square algorithm\\fast-based-correlation feature selection} \\\hline
WSAN				& \makecell[c]{\cite{7913594}\\\cite{8002611}\\\cite{8003415}}			& \makecell[c]{exposition of sensing features\\critical quality
variables estimation\\spatiotemporal feature learning} & \makecell[l]{high-accuracy measurements\\semisupervised deep learning\\deep neural network} \\\hline
NCS					&  \cite{8301555} & cloud virtual machines workload prediction & canonical polyadic decomposition \\\hline
Industrial Robots		&	\makecell[c]{\cite{8024058}\\\cite{8089430}} & \makecell[c]{high-accuracy force tracking in robotized tasks\\feature learning from raw mechanical data} & \makecell[l]{iterative learning with reinforcement\\deep neural network}	\\\hline
Assembly Line			&	\makecell[c]{\cite{7378488} \\\cite{7572996} \\\cite{7999287} \\\cite{8314728}} & \makecell[c]{nonlinear process monitoring\\} & \makecell[l]{optimal operational indices selection\\locally weighted learning\\radial basis function networks\\recursive slow feature analysis}	\\\hline
M2M communication	& \multicolumn{3}{c}{-}	\\\hline
\end{tabular}
\end{center}
\end{table*}

\subsubsection{Job Scheduling}

Job scheduling has been traditionally considered as a core field in the manufacturing research area. The field spans from the single machine scheduling problem which is the simplest type of industrial scheduling problem, to multiple machine scheduling, and even multiple assembly lines scheduling or even inter-factory job scheduling. Examples of single machine scheduling are \cite{doi:10.1080/0951192X.2014.961548}, where nested partitioning-based integration of process planning and scheduling in flexible manufacturing environment is introduced, \cite{Hu2017}, where the authors study the single machine scheduling problem with deadlines where the processing times are described by uncertain variables with known uncertainty distributions, and \cite{doi:10.1080/0951192X.2014.928747}, where the recovery policy of job-shop manufacturing systems is evaluated. Also in \cite{7103301}, the authors propose a software composition method for automated machines that exploits their mechatronic modularity, and they demonstrate that desired behavior of a certain class of machines can be composed of behaviors of its mechatronic components, including fully decentralized scheduling and operation control. Multiple machine job scheduling has been presented in \cite{doi:10.1080/0951192X.2014.964322}, where the authors address the problem of scheduling multi-robot cells with residency constraints and multiple part types, in \cite{7929378}, where the authors consider the serial batching scheduling problem in which a group of machines can process multiple jobs continuously to reduce the processing times of the second and subsequent jobs, and in \cite{Ni2017}, where the authors study a two-machine scheduling problem in fuzzy environments. Multiple assembly lined scheduling is presented in \cite{7842622}, where the authors investigate robust order scheduling problems in the fashion industry by considering the preproduction events and the uncertainties in the daily production quantity. Inter factory scheduling is presented in \cite{doi:10.1080/0951192X.2015.1068450}, where production planning with remanufacturing and back-ordering is discussed, in which there are multiple factories in a cooperative relationship to produce new or remanufactured products.

\subsubsection{Machine Learning}

Machine learning services are by definition data-driven and are used on top of the technological enablers in order to further enhance industrial applications. An outline of the recent industrial machine learning services and the corresponding technical methods used is displayed in Table \ref{tab::ml}. For the IIoT technologies, emphasis has been put on data-driven schemes for predicting the missing QoS values for the IIoT based on kernel least mean square algorithms \cite{LUO201681} and on intelligent IIoT traffic classification using search strategies for fast-based-correlation feature selection \cite{8241382}. WSANs benefit from the exposition of features for sensing that provide high-accuracy measurements for reducing the required manufacturing precision (capacitive displacement sensing in \cite{7913594}). Machine learning is also beneficial for industrial robot enablers, for example with iterative learning procedures with reinforcement for high-accuracy force tracking in robotized tasks \cite{8024058}. Applications in the assembly line focus on process modeling and include data-based methods for automatically selecting optimal operational indices for unit processes in an industrial plant using measured data (without knowing dynamical models of the unit process) \cite{7378488}, data-driven approaches for nonlinear process monitoring under the framework of locally weighted learning \cite{7572996}, using radial basis function networks \cite{7999287}, as well as adaptive process monitoring and fault diagnosis through recursive slow feature analysis \cite{8314728}. Data classification is an active research problem in the industrial data mining and machine learning communities and spreads horizontally over all technological enablers \cite{8006248}. Deep learning, as one of the most important tools of current industrial computational intelligence, achieves high performance in predicting numerous parameters and attributes of industrial applications. However, it is a nontrivial task to train a deep learning model efficiently since the deep learning model often includes a great number of parameters. In \cite{8301555}, the authors introduce an efficient deep learning model to predict cloud virtual machines workload for industrial NCS deployments. In \cite{8002611}, the authors employ deep learning of semisupervised process data with a hierarchical extreme learning machine on a soft sensor industrial application. Spatiotemporal features from sensors can also be learnt through deep neural networks \cite{8003415}. In \cite{8089430}, the authors propose a deep learning network to learn features adaptively from raw mechanical data without prior knowledge.

\subsubsection{Big Data Analytics}

\begin{figure}[t!]
\centering
\includegraphics[width=\columnwidth]{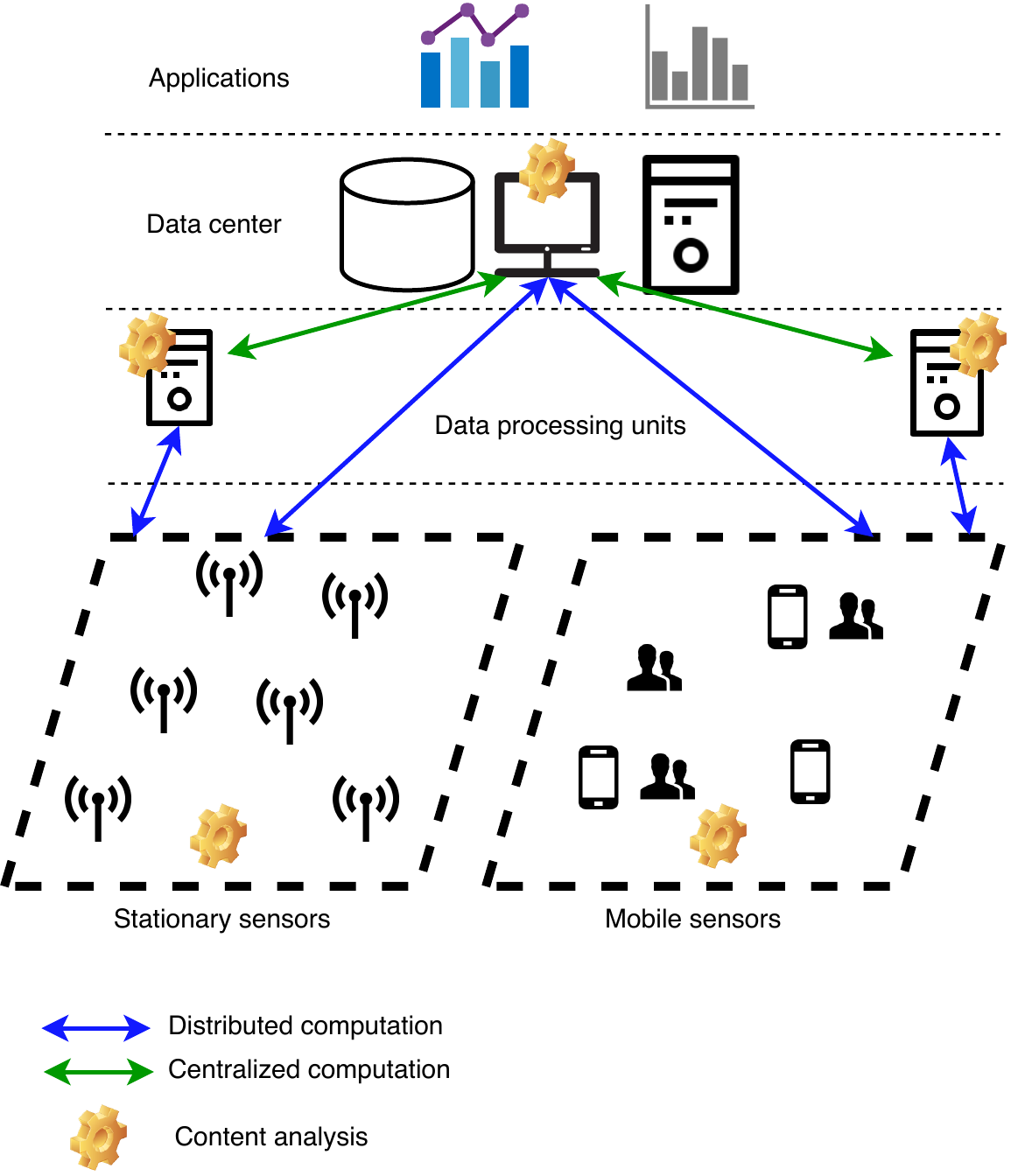}
\caption{Generic big data framework for industrial edge deployments as it is envisioned by recent research approaches.}
\label{fig::bigdata}
\end{figure}

The enormous amount of real-time data is used for the analysis of various industrial applications has led to a trend in I4.0 environments pointing to the use of big-data as a relevant element in the development of next generation industrial systems. Big data analytics offer many opportunities to evaluate data in all layers of the industrial installations, for example, to identify preferences from end-users, to better understand technological enablers' behaviors, or to relate issues derived from a combined and statistical processing of data. The common trend in many current industrial applications is to transfer IIoT data from the physical locations where they are generated to some global cloud platform, where knowledge is extracted from raw data and used to support IIoT applications. Moreover, as \cite{8252775} notes, several big data processes (such as deep learning) require expensive computational resources including high performance computing units and large memory to train a deep computation model with a large number of parameters, limiting its effectiveness and efficiency for industry informatics big data feature learning. Consequently, real-time delay constraints might require that data elaboration or storage is performed at the edge, i.e., close to where it is needed, rather than in remote data centers. However, there are concerns whether this approach will be sustainable in the long run. For this reason, decentralized generic big data framework for industrial edge deployments like the one displayed in Fig.~\ref{fig::bigdata}, as they is envisioned in recent approaches, such as \cite{8231148}, \cite{7857790} and \cite{7835187}, are becoming more and more common. It is visible that the I4.0 trends push towards computation decentralization mainly from the standpoint of data ownership, as well as wireless network capacity. 

Some representative examples of this computation decentralization and of maintaining the data at the edge for distributed operations are the following. In \cite{7111303}, the authors design and test a real-time big data gathering algorithm based on indoor WSANs for risk analysis of industrial operations. In \cite{doi:10.1515/auto-2016-0022}, the authors show different approaches that a classical manufacturing systems company can take into account when applying data mining techniques to address the requirements which come with the IIoT technological enabler. In \cite{7835187}, a distributed  and parallel big data analytics system for modeling and monitoring large-scale plant-wide processes is introduced. In \cite{8047971}, the authors explore the development of an industrial big data implementation able to improve computing performance by splitting the analytic into different segments that may be processed by the engine in parallel using a hierarchical model. Of course, there are also hybrid big data approaches which employ two kinds of computation and data communication: both localized real-time processing and global offline computations. In \cite{7857790}, a manufacturing big data solution for active preventive maintenance in manufacturing environments is implemented. Another hybrid approach is \cite{8291112} which introduces a concentric computing model paradigm composed of sensing systems, outer and inner gateway processors, and central processors for the deployment of big data analytics applications in IIoT. In \cite{8231148}, the authors analyze the relationship between the data processing and the energy consumption through investigating the content correlation of the captured data. Traditional centralized approaches are presented in \cite{FLATH201816}, where the authors develop a big data toolbox for manufacturing prediction tasks to bridge the gap between machine learning research and concrete industrial requirements, and in \cite{Tao2018}, where the authors use big data services in order to design a new method for product design, manufacturing, and service driven by digital twin. Table \ref{tab::bigdata} displays the extent of centrality that the various recent approaches have adopted, in terms of computation for big data analytics.

%\multicolumn{1}{c}{}

\begin{table}[t!]
\begin{center}
\caption{Types of computation for big data analytics.}
\label{tab::bigdata}
\begin{tabular}{ r | l }\hline
\textbf{\makecell[r]{Computation and \\data analytics}} & \textbf{Articles}\\\hline\hline
\makecell[r]{Concentrated \\(cloud / offline)}	& \cite{FLATH201816, Tao2018, 8252775}	\\\hline
\makecell[r]{Distributed \\(edge / real-time)}	& \cite{7111303, doi:10.1515/auto-2016-0022, 7835187, 8047971}	\\\hline
Hybrid 		& \cite{7857790, 8291112, 8231148}	\\\hline
\end{tabular}
\end{center}
\end{table}

\begin{table*}[t!]
\begin{center}
\caption{Data security services for data enabling technologies.}
\label{tab::security}
\begin{tabular}{ r | c | l}\hline
\textbf{\makecell[r]{Data enabling technology}} & \textbf{Articles} &\textbf{Type of security provisioning}\\\hline\hline
IIoT / ICPS			&\makecell[c]{\cite{7866869}	\\\cite{8270567} \\\cite{URQUHART2018450}\\\cite{LIN201842}\\\cite{7927473}}		& \makecell[l]{covert attack for service degradation \\quantification of the impact of cyberattacks on the physical part\\ legal aspects\\blockchain-based remote user authentication with fine-grained access control\\certificateless searchable public key encryption with multiple keywords}	\\\hline
WSAN				& \cite{7029608}			& intercept behavior in the presence of an eavesdropping attacker	\\\hline
NCS					& \makecell[c]{\cite{7090973} \\\cite{7506334}\\\cite{8093698}}	& \makecell[l]{energy efficient intrusion detection\\lightweight secure authentication mechanism for broadcast mode communication\\dynamic cybersecurity risk assessment} \\\hline
Industrial Robots		&	\multicolumn{2}{c}{-}	\\\hline
Assembly Line			&	\multicolumn{2}{c}{-}	\\\hline
M2M communication	& \makecell[c]{\cite{8283561} \\\cite{8291109}}	& \makecell[l]{application-layer traffic filtering\\sensor-cloud trust-based communication}	\\\hline
\end{tabular}
\end{center}
\end{table*}

\subsubsection{Ontologies / Semantics}

In industrial automation, ontology services encompass a representation, formal naming, and definition of the categories, properties, and relations between the data and entities that substantiate various industrial processes. This will lead to the further automation of many tasks in the life cycle of the industrial systems from design to commissioning and operation \cite{8267108}. Those services frequently rely on synergies of industrial standards, such as IEC 61850 \cite{7762918} and IEC 61499 \cite{7569078}, which are used to represent specifications and resulting software models. Due to the fact that semantic data modeling usually deals with data irregularity and diversity, sophisticated dynamic modeling methods have been derived \cite{7917368}. With regards to IIoT and ICPS, OPC-UA and semantic web technologies are able to achieve integration at various levels \cite{8263185}. UML-based approaches can fully automate the generation process of the IIoT-compliant layer that is required for the cyber-physical components to be effectively integrated in the shop-floor \cite{THRAMBOULIDIS2016259}. In order to achieve rapid response to changes from both high-level control systems and plant environment, self-manageable ontological agents can improve flexibility and interoperability \cite{7523919} and automate the process engineering using a knowledge-based assistance system \cite{8290684}. IIoT gateways have already been integrated with dynamic and flexible rule-based control strategies \cite{8091285}. Model-driven NCS enable increased usability \cite{6873317} and model checking \cite{7295624}. In the assembly line, knowledge based ontology services can assist complementary content customization \cite{doi:10.1080/0951192X.2014.880809}, mechanical design knowledge \cite{doi:10.1080/0951192X.2013.874593}, and semantic web service composition \cite{PUTTONEN20151041}. Recognition, semantic annotation and calculating the spatial relationships of a factory's digital facilities \cite{PEREZGALLARDO201640}, as well as the model based synthesis of its automation functionalities \cite{doi:10.1515/auto-2015-0094} are other emerging topics of interest. Ontology services also come handy in cloud manufacturing and take advantage of semantic links to enable automated integrating and distributed updating in resource service clouds \cite{7968321}. Ontology services can also support the development of global production network systems \cite{PALMER201648} and business integration \cite{LI201810} in a more general sense, as well as CAD assembly model retrieval (using multi-source semantics information and weighted bipartite graph \cite{HAN201854}) and visual exploration systems \cite{7390286}.

\subsubsection{Human-in-the-loop}

Human-in-the-loop services, will be an indispensable component of most I4.0 approaches and applications related to the large scale ICPS and assembly line networked environments. This is because large and complex industrial environments necessitate advanced planning and scheduling, careful coordination, efficient communication and reliable activity monitoring, ingredients essential for productivity and safety purposes. A notable relevant area of interest to the researchers recently is human tracking and localization in the industrial facilities. There is a diverse variety of approaches in this field, in terms of generated and used volumes of data. In \cite{LIN2016113}, the authors propose an approach that leverages the inertial sensors embedded in smartphones, uses WiFi fingerprints based on the angle-of-arrival and exploits the ubiquitous presence of diverse data to assist in human localization, thus utilizing data of small volumes. Similarly, in \cite{7733160}, the authors propose a real-time system for human body motion sensing with special focus on joint body localization and fall detection. The proposed system continuously monitors and processes ambient data propagated by industry-compliant radio devices through supporting M2M communication functions. In \cite{7576696}, the authors propose a positioning system for tracking people in highly dynamic industrial environments, such as construction sites. The proposed system leverages the existing CCTV camera infrastructure installed in the industrial environment, along with radio and inertial sensors within each worker's smartphone to accurately track multiple people. Consequently, in this case the data's volume varies according to the data generation source. Even larger volumes of data are used in \cite{8241351}, where the authors employ video analytics in order to implement motion detection framework through motion blobs and successfully provide a features-based person tracking system. Other human-in-the-loop concepts are mobile apps developed to support the customer integration in the product design phase and subsequently the design of the manufacturing network \cite{doi:10.1080/0951192X.2016.1187295}, cross-disciplinary mobile crowdsensing of pervasive sensor data applied in industrial processes \cite{7837601}, as well as automated methodologies for worker path generation and safety assessment \cite{7790844}.

\subsubsection{Security}

Security aspects in factory automation and industrial operations have become a hot topic in the last years since monitoring and control tasks are more and more complex. Also, ICPS are vulnerable to external attacks due to the tight integration of cyber and physical parts. In fact, security incidents such as targeted distributed denial of service (DDoS) attacks on power grids and hacking of factory NCS are on the increase \cite{URQUHART2018450}. Data management in such systems is crucial, as the increased scalability of the deployments can frustrate effective management of security risks, partly due to the complexity of managing the large volumes of data and risks manifesting across interdependent systems. Security has been recently studied across most of the technological enablers presented in this article. Table \ref{tab::security} displays the services that have been presented for security provisioning across the different technologies. In \cite{7866869}, a covert attack for service degradation of ICPS is proposed, which is planned based on the intelligence gathered by another system identification attack. In \cite{8270567}, a risk assessment method is presented targeting the quantification of the impact of cyberattacks on the physical part of ICPS. The proposed method helps carry out appropriate attack mitigation measures. In \cite{LIN201842}, the authors establish a secure remote user authentication with fine-grained access control for IIoT, by proposing a  blockchain-based framework. The proposed framework leverages the underpinning characteristics of blockchain as well as several cryptographic materials to realize a decentralized, privacy-preserving solution. In \cite{7927473}, the authors design a secure channel-free certificateless searchable public key encryption with multiple keywords scheme for IIoT. In \cite{7029608}, the authors study the intercept behavior of an industrial WSAN consisting of a sink node and multiple sensors in the presence of an eavesdropping attacker, where the sensors wirelessly transmit their sensed data. In \cite{7090973}, the authors present an energy efficient intrusion detection and mitigation system for NCS security. The system is data oriented in the sense that it employs data-based selective encryption to reduce energy consumption, and to detect when an attack starts and ends. In \cite{7506334}, the authors present a lightweight secure authentication mechanism for broadcast mode communication in NCS. In \cite{8093698}, a fuzzy probability bayesian network approach for dynamic cybersecurity risk assessment in NCS is proposed. In \cite{8283561}, the authors present a performance model for industrial M2M communication, able to perform advanced application-layer filtering of traffic generated by protocols widely used in industrial deployments (Modbus/TCP). In \cite{8291109}, the authors investigate trust-based communication for industrial deployments, devoting attention to sensor-cloud communication. They propose three types of trust-based M2M communication mechanisms for sensor-cloud. Furthermore, with numerical results, they show that trust-based communication can greatly enhance the performance of sensor-cloud.

\subsubsection{Energy Management}

Energy management for the IIoT and WSANs has naturally received significant attention, as in many cases the devices operate on limited battery supplies (Table \ref{tab::energy}). On the IIoT part, there have been energy efficient improvements on QoS-aware services composition \cite{7452389} (similarly for the ICPS \cite{7898479}), robust authentication protocols \cite{8241345}, routing and data collection \cite{doi:10.1080/0951192X.2017.1285429, 8338161}, as well as resource allocation and utilization \cite{8272309} (similarly for the ICPS \cite{HAN2018205}). On the backbone of the IIoT networks, in the cases where Ethernet is used as an enabler, energy efficiency has also been a timely topic \cite{7588226}. Specifically, in \cite{6787044}, the authors investigate the IEEE 802.3az amendment, known as Energy Efficient Ethernet (EEE) and address its application to Real-Time Ethernet (RTE) networks in factory automation. Additionally, in \cite{7097020}, the same authors expose some data service aspects of the EEE/RTE interplay. 

On the WSAN part energy efficiency is focused on specific data intensive operations. Industrial low power WSAN protocols are one of the key enablers of that revolution but still energy consumption is what is limiting ubiquitous deployments of perpetual and unattended devices \cite{7283537}. Real-time usage data as well as historical data can help identify whether various WSAN components are functioning properly \cite{7403920}. Routing and data collection is traditionally assisted energetically, either through joint data transmission and wireless charging \cite{HAN201619}, or through adjustable data sampling rates \cite{8115313} and distributed and collaborative sleep scheduling \cite{7588228}. Other energy efficient approaches include integrity check in the network \cite{7208869}, node localization \cite{7560611}, data loss minimization \cite{7463484}, and connected target coverage \cite{7369957}. Energy efficient approaches for WSANs of particular interest with respect to the data management mechanisms employed are the following: In \cite{7588229}, the authors apply compressed sensing in order to break the redundant data collection (and thus save significant amounts of energy), by differentiating the available sensed data in principal and redundant, through an online learning component and a local control component. In \cite{7588224}, the authors derive both global and local data storing in the WSAN, and expose the inherent difficulties of each case (data importance degrees definition and data stream reading ability). 

Energy optimization of industrial robotic cells and assembly lines is also essential for sustainable production in the long term. A holistic approach that considers a robotic cell as a whole toward minimizing energy consumption is proposed in \cite{7738397}. Dynamic low-power reconfiguration \cite{6784140} and machine energy consumption minimization \cite{doi:10.1080/0951192X.2014.914631} are key objectives of novel assembly lines. In \cite{8291114}, the authors discuss how dynamic energy management in manufacturing systems can not only solve the current technical issues in manufacturing, but can also aid in the integration of additional energy equipment into energy systems. The significantly important role of data in this process is demonstrated in \cite{doi:10.1080/0951192X.2016.1185154} where the collected data are shown to improve energy consumption awareness and allows the manufacturing energy management systems to make further analysis and to identify where to take actions in the manufacturing process in order to reduce the energy consumption. There have been several energy management and energy consumption optimization methods for the assembly line in the recent literature, with the most notable focusing on production control \cite{7517231}, forecasting models with neural networks \cite{7938652}, mobile service composition \cite{7140840}, real-time demand bidding \cite{7539665}, ontological modeling \cite{doi:10.1080/0951192X.2017.1322220}, process parameter modeling \cite{doi:10.1080/0951192X.2017.1407875}, machine energy consumption profiling \cite{doi:10.1080/0951192X.2017.1339914}, and concurrent energy data collection \cite{doi:10.1080/0951192X.2017.1305508}.

Methodologies and a models which reliably dimension energy scavenger properties to M2M communication requirements and network needs, allowing industries to optimize the adoption of that technologies while keeping technical risks low \cite{6922531}. MAC layer power management schemes which achieves the user specified reliability with minimal power consumption at the node are also of interest to the M2M communication community \cite{8085176}. Interestingly enough, there no significant contributions on energy management issues have been found for the data enabling technology of NCS.

\begin{table}[t!]
\begin{center}
\caption{Energy management for data enabling technologies.}
\label{tab::energy}
\begin{tabular}{ r | l }\hline
\textbf{\makecell[r]{Data enabling technology}} & \textbf{Articles on energy management}\\\hline\hline
IIoT / ICPS	&	 \makecell[l]{\cite{7452389, 7898479, 8241345, doi:10.1080/0951192X.2017.1285429}, \\\cite{8338161, 8272309, HAN2018205, 7588226, 6787044, 7097020}} \\\hline
WSAN	&  \makecell[l]{\cite{7283537, HAN201619, 8115313, 7588228, 7588229}, \\\cite{7208869, 7588224, 7560611, 7463484, 7369957, 7403920}}	\\\hline
NCS 		& - 	\\\hline
Industrial Robots 		&  \cite{7738397}	\\\hline
Assembly Line 		&   \makecell[l]{\cite{6784140, doi:10.1080/0951192X.2014.914631, 8291114, doi:10.1080/0951192X.2016.1185154, 7517231, 7938652}, \\\cite{7140840, 7539665, doi:10.1080/0951192X.2017.1322220, doi:10.1080/0951192X.2017.1407875, doi:10.1080/0951192X.2017.1339914, doi:10.1080/0951192X.2017.1305508}}	\\\hline
M2M Communication 		&  \cite{6922531, 8085176}	\\\hline
\end{tabular}
\end{center}
\end{table}

\subsubsection{Cloud}

Cloud manufacturing has lately gained a fair share of attention from the automation and manufacturing communities. Cloud manufacturing transforms manufacturing resources, capabilities and data into manufacturing services, which can be managed and operated in an intelligent and unified way to enable the full sharing and circulating of manufacturing resources and manufacturing capabilities. Cloud services in the supply chain can greatly reduce time and costs incurred in deploying automation systems, which are quite complex and require large human effort to build \cite{6908023}. Cloud manufacturing can be divided into two categories. The first category concerns deploying manufacturing software on local or global clouds, i.e., a ``manufacturing version'' of cloud computing. The second category has a broader scope, cutting across production, management, design and engineering abilities in a manufacturing business. Unlike with classic computing and data storage, manufacturing involves physical equipment, monitors, materials and so on. In this kind of cloud manufacturing, both material and non-material facilities are implemented on the cloud, in order to support the whole supply chain. The great majority of recent works can be classified in the first category. Cloud manufacturing solutions can be categorized according to the locality of the cloud. In the vast majority of the recent literature the cloud infrastructure is centrally placed, with large public clouds delivering data usually over the internet. In Table \ref{tab::cloud}, the types of data sources and cloud locality in cloud manufacturing are displayed.

\begin{table}[t!]
\begin{center}
\caption{Types of data sources and cloud locality in cloud manufacturing.}
\label{tab::cloud}
\begin{tabular}{ r | c | l | l}\hline
\textbf{\makecell[r]{\makecell[r]{Data enabling \\technology}}} & \textbf{Articles} &\textbf{Data source} & \textbf{Cloud}\\\hline\hline
\makecell[c]{Assembly Line\\Industrial Robots} & \makecell[c]{\cite{7335620} \\\cite{doi:10.1080/0951192X.2015.1067916} \\\cite{doi:10.1080/0951192X.2017.1407446}\\\cite{LIU201472}} & manufacturing resources & \multirow{5}*{global} \\\cline{1-3}
Assembly Line &  \cite{doi:10.1080/0951192X.2017.1314015} & manufacturing services &  \\\cline{1-3}
Assembly Line & \cite{doi:10.1080/0951192X.2015.1067912} & shared memories &  \\\cline{1-3}
WSAN & \cite{LI2017133} & mobile network nodes &  \\\cline{1-3}
NCS & \cite{8252786} & network services & \\\hline
NCS & \cite{7415937} & virtual resources & hybrid \\\hline
IIoT & \cite{8110714} & network devices & local \\\hline
\end{tabular}
\end{center}
\end{table}

As shown in the table, a large portion of works employ global clouds. In \cite{7335620}, the authors target manufacturing resource composition and propose an approach that can better cope with the temporal relationship between the resource services in a business process. In \cite{doi:10.1080/0951192X.2015.1067916}, the authors design a cloud resource sharing based on the Gale-Shapley algorithm and analyze it in the context of fluctuating resource supply and demand. In \cite{doi:10.1080/0951192X.2017.1407446}, the authors present an agent-adapter-based method of for manufacturing clouds to enable manufacturing with various physically connected machines from geographically distributed locations over the Internet. In \cite{LIU201472}, the authors suggest a multi-granularity resource virtualization and sharing method for cloud manufacturing. In \cite{doi:10.1080/0951192X.2017.1314015}, the authors introduce service clustering network-based service composition. In this approach, services are first clustered into abstract services, and then a clustering network of the abstract services is established. In \cite{doi:10.1080/0951192X.2015.1067912}, the authors design an effective load-adjusted allocation algorithm for enhancing memory reusability and improving the performance of servers by balancing their workloads. In \cite{LI2017133}, the authors consider industrial WSAN with mobile nodes and propose a fixed-path mobile node handover strategy, assisted by cloud services and an ants-colony algorithm. In \cite{8252786}, the authors propose a cloud-based decision support system for self-healing in distributed automation systems using fault tree analysis. Some fewer recent works employ hybrid or local clouds. In \cite{7415937}, the authors study the problem of how to maximize the profit of a local (private) cloud in architectures of a combination of local and global (hybrid) clouds while guaranteeing the service delay bound of delay-tolerant tasks. In \cite{8110714}, the authors suggest an embedded cloud database service method for distributed IIoT monitoring.

\section{Open Research Challenges} \label{sec::open}

In this section, we identify some open research challenges on data management in industrial networked environments and their inherent tradeoffs. Subsequently, we focus our attention on a wide variety of thematic topics pertaining to the requirements of data management, as presented in the previous sections. These notes provide crisp insights for the design of future data management applications.

\subsection{Energy efficient data delivery with small delays}

\begin{figure}[t!]
\centering
\includegraphics[width=\columnwidth]{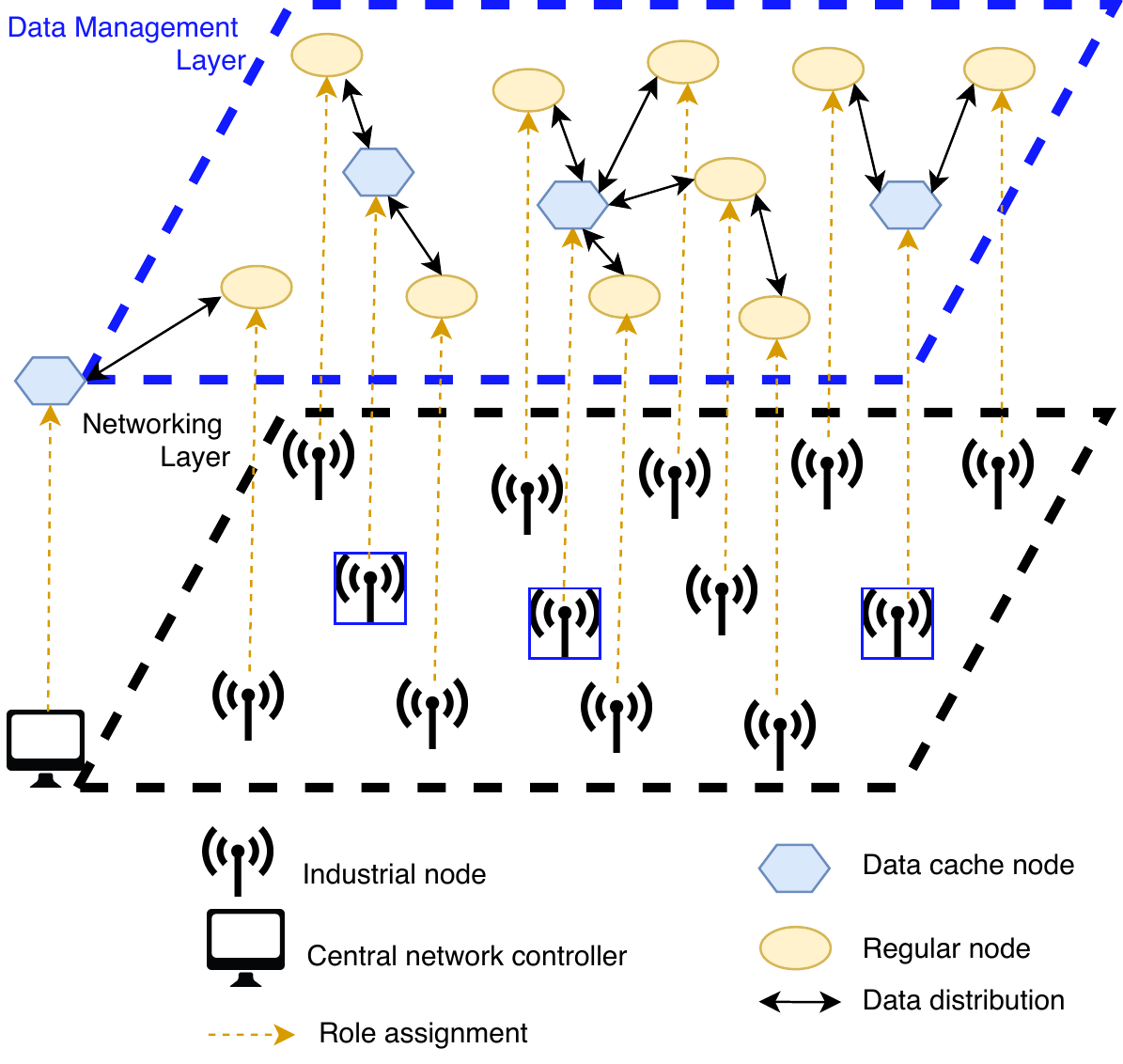}
\caption{Conceptual design of a data management layer over an industrial network.}
\label{fig::dml}
\end{figure}

Ensuring energy efficient, low-latency data delivery in industrial networked environments is of capital importance and is currently receiving more and more attention in academia and industry. However, in current industrial configurations, the computation of the data exchange and distribution schedules is quite primitive and highly centralized. Usually, the generated data are transferred to a central network controller using wireless or wired links. The controller analyzes the received information and, if needed, reconfigures the network paths and the data forwarding mechanisms, and changes the behavior of the physical environment through actuator devices. Traditional data distribution schemes are usually implemented over relevant industrial protocols and standards, like WirelessHART, 802.15.4e and 6TiSCH. Those entirely centralized and offline computations regarding data distribution scheduling, can become inefficient in terms of end to end latency. Additionally, in industrial environments, the topology and connectivity of the network may vary due to link and sensor-node failures. Also, very dynamic conditions, which make communication performance much different from when the central schedule was computed, possibly causing sub-optimal performance, may result in not guaranteeing energy requirements. These dynamic network topologies may cause a portion of industrial nodes to malfunction. With the increasing number of involved battery-powered devices, industrial networks may consume substantial amounts of energy; more than needed if local, distributed computations were used. In order to address those emerging challenges of the I4.0, novel data management layers have to be engineered over the device and networking planes of the industrial deployments. Those layers have to operate independently from and to complement the routing process, targeting at distributing the data in the networks in a decentralized manner, while at the same time respecting the strict I4.0 requirements. In fact, not all data need to be transferred to central network controllers prior to delivery to the data consumers (as traditional industrial routing approaches usually impose); in fact, data can be also stored managed locally at selected data cache nodes (Fig.~\ref{fig::dml}), exploiting, when needed, additional levels of information. 

%\subsection{Distributed data processing}

\subsection{Data distribution in local and mobile clouds}

\begin{figure}[t!]
\centering
\includegraphics[width=\columnwidth]{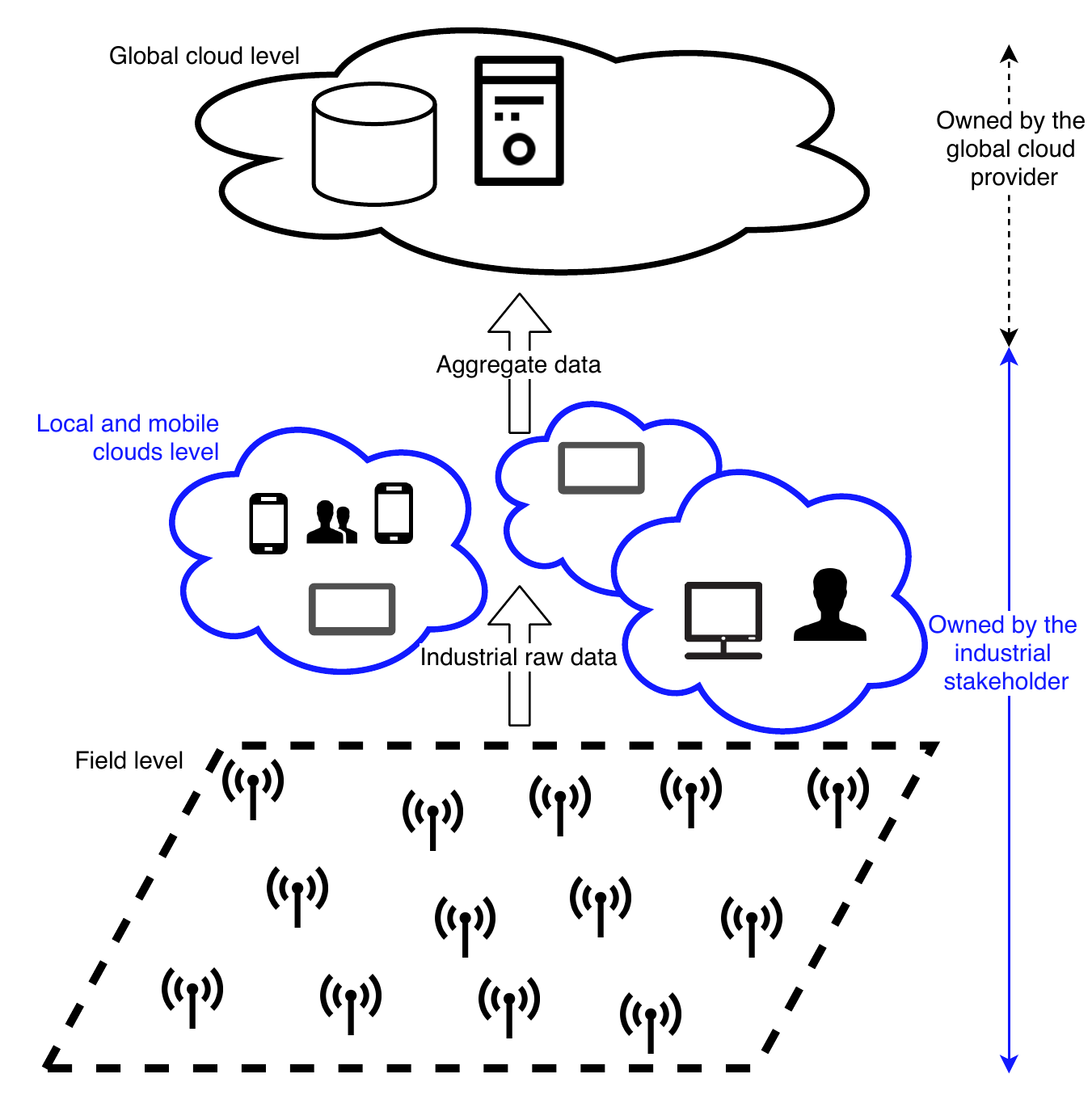}
\caption{Conceptual design of a multi-layer cloud platform.}
\label{fig::cloud}
\end{figure}

As shown in Table \ref{tab::cloud} the most common current approach for collecting and processing large volumes of data for cloud manufacturing purposes is based on the assumption that some network infrastructure is able to support the collection and delivery of all these data toward the cloud, which is intended to be the back-end aimed at processing and getting value from such data. In general IIoT/ICPS environments, this backbone is a wideband cellular network such as LTE. In the case of manufacturing environments this may also be the case, or more localized wideband infrastructures such as WiFi may be used. In any case, an approach relying exclusively on global cloud providers to provide holistic industrial data services has limitations from two main standpoints. On the one hand, wideband wireless networks may not provide sufficient bandwidth so support the data traffic demand. On the other hand, relying only on global clouds deployed may make manufacturing stakeholders to loose control on their data, as data will be transferred to data centers without any control of the data owner. In addition, meeting the manufacturing stakeholders requirements in terms of storage and computation capacity may have a significant impact on the cost incurred by the stakeholders for ICT services, which, if reduced, could be more profitably invested in the core production process. In order to overcome these issues there is a need of a paradigm shift in the way the gathered data is managed and processed. To this end, the employment of local and mobile cloud technologies as a way to implement a multi-layer cloud infrastructure would be necessary (Fig.~\ref{fig::cloud}). This will enable the exploitation of not only global cloud services, but also local resources available at the stationary and mobile devices of the industrial deployments. In such environment, a number of mobile devices (e.g. the devices of various operators working at the manufacturing premises) are available, and typically their computation and storage resources are underutilized. Instead of relying exclusively on storage and computation services provided by a global cloud provider, the storage and the computation tasks can be distributed among those local devices, that will therefore form a local (and in some cases mobile) cloud. In this paradigm, global cloud services can be used only when (i) global information is needed in order to better analyze the status of the production process, or (ii) local resources are saturated and additional capacity is needed. For example, storage available at local devices would be enough only for storing information about parts produced in a limited time window in the past. Older data may be stored on a global cloud storage service, possibly in an encrypted form. However, data related to most recently produced parts would still be available locally, and could be accessed without transferring back and forth them between local devices and global cloud data centers. The resulting solution will be a multi-layer cloud platform, whereby global resources and local resources will be used elastically and in a synergic way, depending on the need of the virtual metrology service.

%\subsection{Data privacy}

\subsection{Distributed, real-time data security for industrial robots and assembly line}

As shown in Table \ref{tab::security}, there is a lot of work already implemented in terms of data security for IIoT/ICPS, WSANs, NCS and M2M Communication. However, the absence of security mechanisms for the technological enablers of the assembly line and the industrial robots is notable. More than that, the decentralization of the production process, the integration with IIoT technologies (the nature of which makes them vulnerable) and the introduction of open and ubiquitous data, leaves the assembly lines and robots further exposed to external threats. To date, security has not been a concern for the (in many cases legacy) assembly lines and industrial robots. Yet, practitioners have recognized that the open and uncontrollable nature of the M2M communication enabler opens these systems to a variety of possible security threats and vulnerabilities. Security solutions will also need to be operated in a distributed manner, because centralized solutions require transmitting data to the central controller, which may result in data loss and delay to the threat detection decisions, particularly in large-scale deployments. In contrast, distributed solutions are much more agile and robust to data transmission failures and, more importantly, scale to larger sizes. For example, industrial anomaly detection for malicious attacks (e.g., false data injection) can be performed either at the central controller or at local distributed devices \cite{6882174}. Finally, following the same example, since real-time information is critical
and even a single abnormal security behavior may lead to a catastrophic cascade of failures throughout the whole system,  abnormalities should be detected as early as possible to minimize the possibility of potential damage. To achieve this, real-time data security solutions will be able to provide online threat detection is needed. Those solutions should be able to identify the anomaly condition of each observation, as soon as the local data observations are collected.

\subsection{Convergence between industrial / automation / manufacturing and communication / networking / computation} \label{sec::convergence}

NCS currently provide deterministic services for the assembly line and the industrial robots, while the IIoT and the WSANs provide best effort services for the entire automation pyramid. Also, as it was demonstrated in Table \ref{tab::arch}, the recent architectural trends for assembly line and industrial robot installments are focusing on centralized data management, while the trends for IIoT and WSANs are pushing towards decentralization, mostly due to the emerging data ubiquity. It has already been argued that a convergence should occur, and that future converged industrial deployments should support both best effort and deterministic services, with very low latency and jitter \cite{7498096}. This convergence is motivated even more and will be further extended with the pervasiveness and the variety of different data sources in the shop-floor. Consequently, industrial automation providers face a challenge and can significantly benefit from communication/networking technologies and services. If they are not able to find powerful and flexible computing services that would enable them to store and process ``as required'' the manufacturing information they have generated, they will never be able to leverage on faster and more complete control of the production process in the digital domain to gain a competitive advantage. If they remain to perform the analysis as they currently have to perform, i.e. on the physical domain, they will continue suffering a negative impact on production yield and costs.

\section{Conclusions}
In this survey article we reviewed the recent literature (2015-2018) on data management as it applies to networked industrial environments. Of particular interest to our review have been the data enabling technologies and the data centric services that both the Communications/Networking/Computation field and the Industrial/Manufacturing/Automation field are providing, in order to boost the production performance and address the emerging I4.0 requirements. We focused the survey at first on recent practical use cases and emerging architectural trends, where we made a note on the convergence that should occur between the two scientific fields, so as to enable an efficient future data management approach. Then, we performed an exhaustive survey on the most relevant and acclaimed research journals and came up with a taxonomy of the recent works in technologies and services. Finally, after this holistic research, we identified several interesting open challenges for the future; energy efficient data delivery with small delays, data distribution in local and mobile clouds, distributed, real-time data security for industrial robots and assembly line, and convergence between the two main scientific fields.

%\section*{Acknowledgment}
%This work has been funded by the European Commission through the FoF-RIA Project AUTOWARE: Wireless Autonomous, Reliable and Resilient Production Operation Architecture for Cognitive Manufacturing (No. 723909).

\ifCLASSOPTIONcaptionsoff
  \newpage
\fi



%\bibliographystyle{IEEEtran}
%\bibliography{data-bib}


% \end{thebibliography}

\balance


\end{document}